\documentclass{article}
\usepackage{arxiv}

\usepackage{graphicx,color,amsmath,amsfonts,amscd,amssymb,pstricks,tikz}
\usepackage{dcolumn}
\usepackage{bm}
\usepackage{longtable}
\usepackage{amsmath}
\usepackage{mathrsfs}
\usepackage{amsfonts}
\usepackage{textcomp}
\usepackage{esvect}
\usepackage{float}
\usepackage[USenglish,british]{babel}
\usepackage{environ}
\usepackage{xifthen}
\usepackage{xargs}			
\usepackage{hyperref}
\usepackage{physics}
\usepackage{multirow,here}

\newcommand{\lc}{L_{\rm c} }
\newcommand{\lb}{L_{\rm b} }

\newcommand{\lqa}{L_{\rm q} }
\newcommand{\alp}{\alpha_{\rm c} }
\newcommand{\lcs}{L^2_{\rm c} }
\newcommand{\muc}{\mu_{\rm c} }
\newcommand{\much}{\displaystyle{\frac{\mu_{\rm c}}{2}}}
\newcommand{\mucp}{\mu'_{\rm c} }
\newcommand{\phc}{\phi_{\rm c} }
\newcommand{\phcs}{\phi^2_{\rm c} }
\newcommand{\rc}{\rho_{\rm c} }
\newcommand{\bma}{\beta_{\rm max} }
\newcommand{\bmi}{\beta_{\rm min} }
\newcommand{\dma}{D_{\rm max} }
\newcommand{\dmi}{D_{\rm min} }

\newcommand{\ksf}{K_{3\rm F} }
\newcommand{\ksd}{K_{3\rm D} }
\newcommand{\laf}{\lambda_{\rm F} }
\newcommand{\lad}{\lambda_{\rm D} }
\newcommand{\lafh}{\frac{\lambda_{\rm F}}{2} }
\newcommand{\ladh}{\frac{\lambda_{\rm D}}{2} }
\newcommand{\laff}{\frac{\lambda_{\rm F}}{4} }

\newcommand{\gtr}{\gamma_{\rm tr}}
\newcommand{\ie}{i.e.\ }

\newcommand{\eg}{e.g.\ }

\title{Combined-function optics for circular high-energy  hadron colliders}
\author{M. Giovannozzi\thanks{Corresponding author: massimo.giovannozzi@cern.ch} \\
Beams Department, CERN, Esplanade des Particules 1, 1211 Meyrin, Switzerland \\
\And
E.~Todesco \\
Technology Department,  CERN, Esplanade des Particules 1, 1211 Meyrin, Switzerland}
\begin{document}
\maketitle

\begin{abstract}%
The design of future circular high-energy hadron colliders is based on the achievement of challenging magnetic fields, needed to keep the hadron beams orbiting along the ring circumference. The strength of the dipolar magnetic field is a function of the machine radius, beam energy, and of the fraction of the ring circumference that can be filled with dipoles. In this paper we propose to use a combined-function periodic cell to maximise the filling factor of the dipole magnets. The optical properties of the proposed periodic structure are discussed in detail together with the design of the superconducting magnets needed to implement the proposed approach.
\end{abstract} 
%
\section{Introduction}
The advent of the alternating gradient (AG) paradigm, also called strong focusing, revolutionised the domain of accelerator physics and technology~\cite{doi:10.1119/1.1941877}. In the seminal papers~\cite{Christofilos,PhysRev.88.1190,PhysRev.91.202.2,Courant:593259} are laid the foundations of the modern field of accelerator physics. The novel principle of AG machines allowed a breakthrough in the performance of circular particle accelerators, which is testified by the jump in beam energy, from $\approx 3$~GeV of the Cosmotron~\cite{Blewett1953TheCR} to $\approx 30$~GeV, of the first two rings designed based on this principle, namely the CERN Proton Synchrotron (PS)~\cite{Regenstreif:214352,Regenstreif:102347,Regenstreif:278715,Regenstreif:342915,Regenstreif:342916,Burnet:1359959}, and the BNL Alternating Gradient Synchrotron (AGS)~\cite{Brown:2264406,Roser,Green}. Then the CERN Intersecting Storage Ring (ISR)~\cite{ISRDR}, the world's first hadron collider, followed the first two AG machines. Since then, several other hadron circular accelerators were built around the world, with even higher beam energies. It is worth noting a major difference between the first two rings and the others, as the normal-conducting magnets were combined-function devices, with dipolar and quadrupolar magnetic fields superimposed. This feature has led to a compact magnet design, thus leading to relatively small machines for the maximum beam energy possible.

The next technological jump was the advent of superconducting magnets, which pushed even further the energy frontier of circular storage rings and colliders (see, \eg \cite{Tollestrup:1141042} and references therein for a full account on the topic). The limit of beam energy is in the process of being increased even more by the CERN study on the Future Circular Collider (FCC), in the form of a hadron collider with $50$~TeV beam energy~\cite{FCC-hhCDR}. 

Recently, it has been suggested to the authors~\cite{devred} to consider whether the quest for high energy could be made more efficient by resuming the old concept of combined-function magnets. This could lead to a couple of important improvements: the removal of the arc quadrupoles, which represents a non-negligible saving in the cost of the accelerator, plus the resources needed for the magnet developments, etc. Today, the nominal layout of the FCC-hh, based on the LHC arc, considers having a group of correctors (orbit dipole correctors, tuning quadrupoles, skew quadrupoles, chromatic sextupoles, Landau octupoles) gathered in the cryostat hosting the cell quadrupole. The filling factor of the accelerator could be further increased if one could spread these correctors in the combined-function dipoles. This possibility will be explored in a forthcoming paper, and this advantage could be used to reduce the dipole field, always at constant beam energy, whose marginal cost is extremely high in the $15-16$~T range. It is also worth mentioning that in the framework of the Muon Accelerator Program (MAP), superconducting combined-function magnets have been studied~\cite{Kashikhin:IPAC12-THPPD036}. However, the design principle is to have nested dipolar and quadrupolar coils (in any of the configurations possible, \ie dipole as inner or outer coils), which is not what is proposed in this article, where a unique set of coils generates a superposition of dipole and quadrupole fields. 

In this paper, the features of a combined-function periodic cell are considered in detail in order to sketch a design of a possible periodic structure for a high-energy collider. The properties of the combined-function superconducting magnets are then studied and a possible design is presented, which complies with the capabilities of the current magnet technology, \ie assuming a peak field of $16$~T, which is the nominal value assumed for the FCC-hh design~\cite{FCC-hhCDR}. It is worth mentioning~\cite{TTandFZ} that a periodic cell based on combined-function magnets had been considered in the early days of the LHC design~\cite{Meot:703530}, but it was not retained. The option of using a combined-function optics for a partial energy upgrade of the CERN LHC ring was mentioned in~\cite{Fartoukh:2293138}, but not evaluated in any detail. On the other hand, in~\cite{PhysRevAccelBeams.23.101602}, in the framework of a full energy upgrade of the LHC, the so-called HE-LHC~\cite{HE-LHCCDR}, a combined-function cell had been considered. However, in that case the term combined-function optics has a different meaning from that used in this paper. The proposed HE-LHC periodic cell had a standard FODO structure \ie with separated-function quadrupoles, but the superconducting dipoles featured a nonzero quadrupolar component. In this paper, we focus on a pure combined-function periodic cell to push for the maximum possible benefits for the ring design and its performance.

The plan of the paper is the following: in Section~\ref{sec:FODO} the basic properties of a FODO cell are recalled, with particular emphasis on the correction of chromatic effects. In Section~\ref{sec:CFoptics} the properties of the optical parameters of a generic combined-function cell are presented and discussed in detail, with Section~\ref{sec:optcomparison} detailing a comparison of the optical parameters of the two cell types, and Section~\ref{sec:chromcorr} discussing in detail the essential topic of chromaticity correction. In Section~\ref{sec:layout} a possible cell layout, based on the combined-function concept, is presented. The properties of the superconducting magnets needed to implement the proposed concept are discussed in Section~\ref{sec:magnets}, and based on this discussion an optimised layout is presented in Section~\ref{sec:opt}. Finally, conclusions are drawn in Section~\ref{sec:conclusions}. A collection of results on the optical properties of a periodic FODO cell is presented in Appendix~\ref{sec:app1}, the impact on the longitudinal dynamics of the optical parameters of the proposed combined-function periodic cell is discussed in Appendix~\ref{sec:app2}, and some considerations on an asymmetric combined-function cell design are presented in Appendix~\ref{sec:app3}.
\section{Generalities on the FODO periodic cell} \label{sec:FODO}
\begin{figure}[htb]
\centering
\begin{tikzpicture}
\draw[thick, dashed] (0.1,0) -- (12.3,0);
\node at (6.2,0.3) {$\lc$};
\draw[thick, dashed] (-1,2) -- (13.4,2);
\draw[thick,->] (0,1) rectangle (1,3);
\node at (0.5,2.25) {$Q_\mathrm{F}$};
\node at (0.5,3.3) {$\lqa/2$};
\draw[thick,->] (1.1,1.5) rectangle (5.1,2.5);
\node at (3.1,3.3) {$\lb/2$};
\node at (3.1,2.25) {$B$};
\draw[thick,->] (5.2,1) rectangle (7.2,3);
\node at (6.2,2.25) {$Q_\mathrm{D}$};
\node at (6.2,3.3) {$\lqa$};
\draw[thick,->] (7.3,1.5) rectangle (11.3,2.5);
\node at (9.3,3.3) {$\lb/2$};
\node at (9.3,2.25) {$B$};
\draw[thick,->] (11.4,1) rectangle (12.4,3);
\node at (11.9,2.25) {$Q_\mathrm{F}$};
\node at (11.9,3.3) {$\lqa/2$};
\draw[thick, dashed] (-1,-1.5) -- (13.4,-1.5);
\draw[thick, dashed] (0,0.9) -- (0,-0.9);
\draw[thick, dashed] (12.4,0.9) -- (12.4,-0.9);
\draw[thick,->] (0,-1.0) rectangle (3.05,-2.0);
\node at (1.525,-0.7) {$\lqa/4=\lb/4$};
\node at (1.525,-1.25) {$B+Q_\mathrm{F}$};
\draw[thick,->] (3.15,-1.0) rectangle (9.25,-2.0);
\node at (6.2,-0.7) {$\lqa/2=\lb/2$};
\node at (6.2,-1.25) {$B+Q_\mathrm{D}$};
\draw[thick,->] (9.35,-1.0) rectangle (12.4,-2.0);
\node at (10.875,-0.7) {$\lqa/4=\lb/4$};
\node at (10.875,-1.25) {$B+Q_\mathrm{F}$};
\end{tikzpicture}
\caption{Top: sketch of the layout of a periodic FODO cell. In the case of the thin-lens approximation, the quadrupoles are replaced by thin elements located at $0, \lc/2, \lc$, whereas the thin dipoles are located at $\lc/4, 3\lc/4$. Bottom: sketch of the layout of a periodic combined-function cell. The type of the magnetic field generated by each element is also shown.}
\label{fig:cell}
\end{figure}
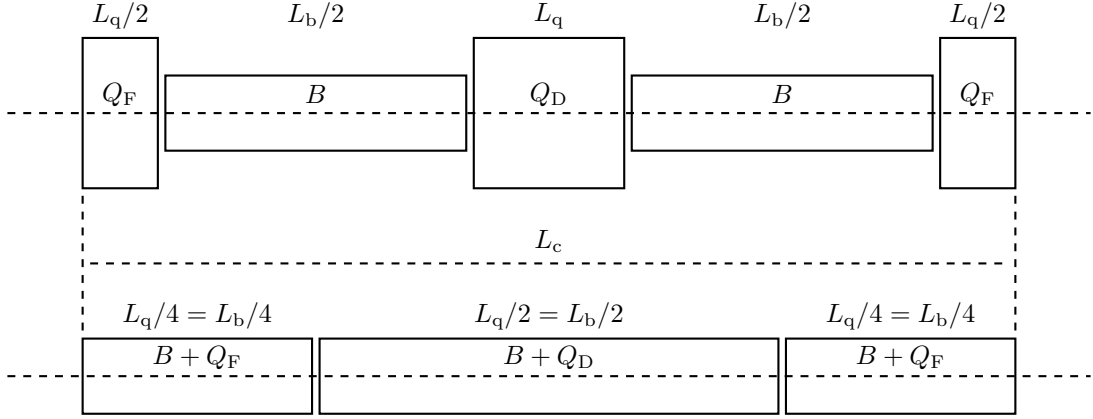

A FODO cell\footnote{This standard acronym stands for Focusing, Off, Defocusing, Off, to indicate that the sequence of quadrupolar fields along the cell are interspersed by regions without any focusing field.} is the standard periodic structure that constitutes the building block of modern circular storage rings and colliders. In its essence it is made of magnets generating dipolar and quadrupolar fields, the dipoles being located in between the two main quadrupoles that feature fields of opposite type, \ie focusing and defocusing to implement the AG principle (see Fig.~\ref{fig:cell}, top, for a sketch of a system of length $\lc$). 

The key properties of this lattice structure have been derived in the seminal paper~\cite{Courant:593259} in which the principles of strong focusing have been studied in detail. The optical proprieties of such a structure can be derived by using the so-called thin-lens approximation, where the quadrupoles are located at positions $0, \lc/2, \lc$, while the dipoles are at positions $\lc/4, 3\lc/4$. The main results for such a structure are collected in Appendix~\ref{sec:app1}. A FODO cell layout is characterised by three parameters: the cell length, $\lc$, the total deflection angle, $\phc$, and the total phase advance, $\muc$. The first two parameters are related to the geometrical layout of the periodic cell, the second parameter is linked to the integrated magnetic strength of the dipoles, while the phase advance is linked to the integrated magnetic strength of the quadrupoles (see Appendix~\ref{sec:app1}) and is related to the optical properties of the cell. 

In terms of optical parameters, the maximum and minimum values of the $\beta$- and dispersion-function are of paramount importance and feature some interesting scaling laws in terms of cell parameters. Both the $\beta$- and dispersion-functions are linear in $\lc$, and the dispersion-function is also linear in $\phc$.

An important concept in the design of a periodic structure and of a circular accelerator, is the dipole filling factor, \ie the fraction of the cell length that is occupied by dipole magnets. This parameter is essential due to its link with the energy reach of the circular particle accelerator, as it provides the value of the integrated dipolar field that can be used to bend the particles and keep them on the reference orbit. If $\lb$ stands for the total length of the dipoles installed in the FODO cell, then the filling factor $\chi$ is given by $\chi=\lb/\lc$, and typical values are $\chi_\mathrm{LHC} \approx \chi_\mathrm{FCC-hh}\approx 0.8$. Note that this definition disregards any engineering detail, such as the need of free longitudinal spaces to ensure the interconnection of the various superconducting magnets installed in the cell. In this sense, $\chi$ is an upper bound on the actual filling factor achievable in the tunnel. The importance of the quantity $\chi$ is that it enters in the fundamental relation
\begin{equation}
    B_\mathrm{max} \, \chi \, \rc = \frac{e}{c} \, p_\mathrm{beam}
\end{equation}
connecting the maximum dipolar field $B_\mathrm{max}$, which is a function of the technology used for the superconducting magnets, the bending radius $\rc$, which is proportional to the dimension of the tunnel, and the beam momentum $p_\mathrm{beam}$, so that $\chi \to 1$ is the goal of any high-energy collider. Such a condition might seem to be achievable by designing very long cells, but this should be balanced by the increase of beam size for such cells. Indeed, starting from Eq.~\eqref{eq:grad_mu} of the Appendix~\ref{sec:app1} and the expression of $\bma$ it is possible to show that 
\begin{equation}
    L_\mathrm{q, min}=\frac{\sqrt{2 \, \epsilon_\mathrm{N}}\, N_\sigma}{\eta \, \phc} \sqrt{\tan \much \left ( 1+ \sin \much \right )} \sqrt{\lc} \propto \sqrt{\lc} \, ,
\end{equation}
where the maximum gradient used to minimise the quadrupole length is given by $G_\mathrm{max} \, r_\mathrm{coil}=B_\mathrm{max}$, $r_\mathrm{coil}$ being the radius of the coils. The requested beam aperture is $A=\eta \, r_\mathrm{coil}$, which is linked to the beam size by $A=N_\sigma \sqrt{\epsilon_\mathrm{N}\bma}$, in which the contribution of the dispersion to the betatronic beam size has been neglected and $N_\sigma$ stands for the number of beam sigma that should fit inside the aperture $A$. $\eta$ represents a design parameter defining the fraction of the coil aperture that is available to the beam. Therefore, $\chi -1 \propto \lc^{-1/2}$, but the beam size increases as $\lc^{1/2}$ so that a trade off has to be found between cell length and magnet aperture, which in the end prevents $\chi$ from approaching unity.

Another key feature of a periodic structure is its chromatic properties. These have to do with the off-momentum dynamics of the charged particles, which have to be compensated for by a set of dedicated sextupolar magnets (see, \eg \cite{bryant_johnsen_1993}). At the lowest order, the off-momentum dynamics has an impact on the betatron tunes that are moved from their nominal values due to the so-called chromaticity (see Appendix~\ref{sec:app1}). For this reason, it is customary to locate two independently powered sextupole magnets close to the focusing and defocusing quadrupoles of the FODO cell, respectively. The mathematical detail of the chromaticity correction is reported in the Appendix~\ref{sec:app1}, and here we would like to stress that the required strength of the sextupoles is obtained by solving a linear system of equations that involve the beta- and dispersion-functions of the FODO cell. In particular, given the standard location of the sextupoles, the difference between $\bma$ and $\bmi$ plays an essential role in determining the strength of the sextupoles, in the sense that a reduction of such a difference increases the required strength. 

The chromatic sextupoles are a source of nonlinearity in the beam dynamics, which perturbs the beam behaviour. The global impact of nonlinearities can be estimated by means of the so-called resonance driving terms (RDT) (see~\cite{PhysRevSTAB.3.054001,PhysRevSTAB.8.024001,PhysRevSTAB.10.034002,PhysRevSTAB.14.034002,PhysRevSTAB.17.074001,PhysRevAccelBeams.20.091001} for a selected list of references), which are expressed, for the case of sextupole magnets and in first order with respect to their strength as
\begin{equation}
    \begin{split}
        c_{m,n}(s) & = \frac{1}{2\pi} \int_s^{s+\lc} \dd s' \tilde{K}_{3}(s') \beta_x(s')^{\frac{|m|}{2}} \beta_y(s')^{\frac{|n|}{2}} \exp \left \{ i \left [ m \left ( \mu_x(s')-\mu_x(s) \right ) + n \left ( \mu_y(s')-\mu_y(s) \right )\right ] \right \} \\
        & = \frac{\exp \left [ -i \left ( m \mu_x(s) + n \mu_y(s) \right ) \right ]}{2\pi} \int_s^{s+\lc} \dd s' K_{3}(s') \beta_x(s')^{\frac{|m|}{2}} \beta_y(s')^{\frac{|n|}{2}} \exp \left [ i \left ( m \mu_x(s') + n \mu_y(s') \right ) \right ]
    \end{split}
    \label{eq:rdt}
\end{equation}
where $|m|+|n|=3$ is the order of the resonance under consideration and
\begin{equation}
    \tilde{K}_{3}(s)=\frac{1}{B_0\rc} \frac{\partial^2 B_y(s)}{\partial x^2}
\end{equation}
is the strength of the sextupolar field ($B_0\rc$ being the magnetic rigidity). Under the assumption that the sextupoles are represented as thin lenses, the integral can be replaced by a finite sum, and when the strength of the chromatic sextupoles is connected to the chromaticity value and the FODO cell properties (see, Appendix~\ref{sec:app1}, Eq.~\eqref{eq:chrom_corr3}) with 
\begin{equation}
     K_{3}=\frac{\ell}{B_0\rho} \frac{\partial^2 B_y}{\partial x^2}
\end{equation}
being the integrated sextupolar strength ($\ell$ being the magnetic length of the sextupoles), and one obtains the following expressions for the main sextupolar RDTs~\cite{Fartoukh:1414744} 
\begin{equation}
    \begin{split}
        |c_{3,0}(0)|  & = 16 \pi \Delta Q' \frac{ \lc^{-1/2}}{\phc} \frac{\sin^3 \muc}{\sin \much \left ( 4-\sin^2 \muc \right )} \left | \left ( 2-\sin \much \right ) \left ( 1 + \sin \much \right )^{3/2} + \right .\\
        & \left . -  \left ( 2+\sin \much \right ) \left ( 1 - \sin \much \right )^{3/2} \exp \left ( 3 i \much \right ) \right | \\
        & \\
        |c_{1,2}(0)|  & = 16 \pi \Delta Q' \frac{ \lc^{-1/2}}{\phc} \frac{\sin^3 \muc}{\sin \much \left ( 4-\sin^2 \muc \right )} \left | \left ( 2-\sin \much \right ) \left ( 1 + \sin \much \right )^{1/2} \left ( 1 - \sin \much \right ) + \right .\\
        & \left . -  \left ( 2+\sin \much \right ) \left ( 1 - \sin \much \right )^{1/2} \left ( 1 + \sin \much \right ) \exp \left ( 3 i \much \right ) \right | \\
        & \\
        |c_{1,-2}(0)| & = 16 \pi \Delta Q' \frac{ \lc^{-1/2}}{\phc} \frac{\sin^3 \muc}{\sin \much \left ( 4-\sin^2 \muc \right )} \left | \left ( 2-\sin \much \right ) \left ( 1 + \sin \much \right )^{1/2} \left ( 1 - \sin \much \right )^{1/2} + \right .\\
        & \left . -  \left ( 2+\sin \much \right ) \left ( 1 - \sin \much \right )^{1/2} \left ( 1 + \sin \much \right )^{1/2} \exp \left ( - i \much \right ) \right | \, ,
    \end{split}
\end{equation}
where the dependence of the FODO-cell characteristics, namely $\lc, \phc, \muc$, and the chromaticity change $\Delta Q'$ is clearly visible.
\section{The combined-function periodic cell} \label{sec:CFoptics}
\subsection{General properties}
The layout of the proposed combined-function cell is shown in Fig.~\ref{fig:cell} (bottom). It is made of magnets in which the dipolar component is superimposed on a quadrupolar one, either of focusing or defocusing type. For this periodic structure, the thin-lens approximation is not appropriate and will never be used in the following. The structure is fully symmetric with respect to the length of the focusing and defocusing magnets, although it is worth mentioning that asymmetric designs could be envisaged and some of their properties are reported in Appendix~\ref{sec:app3}. The first comment about the proposed layout is that this periodic cell features $\chi=1$, under the same assumptions used for the FODO cell, \ie no space for interconnections is considered, which is an essential advantage with respect to its FODO counterpart. 

Neglecting the weak focusing generated by the dipoles, which is always an excellent approximation given the target application of high-energy rings, the $4\times 4$ cell transfer matrix $M(\laf,\lad)$ can be written, using $\lambda_\mathrm{F,D}=\sqrt{K_\mathrm{F,D}} \lc$, where $K_\mathrm{F,D}$ stand for the normalised quadrupolar gradient, as
\begin{equation}
    M = \left ( 
    \begin{array}{cc}
    M_x &  0\\
    0   & M_y
    \end{array} \right )
\end{equation}
with $M_{z}$ (note that in the paper the variable $z$ is used to indicate generically the variable $x$ or $y$) being $2\times2$ matrices given by 
\begin{equation}
\begin{split}
    M_{x,\mathrm{F}} & = \left(
\begin{array}{cc}
 \frac{\lad^2-\laf^2}{2 \laf \lad}\sinh \ladh \sin \lafh +\cosh \ladh \cos \lafh & -\frac{\lc (\lad^2-\laf^2)}{2 \lad \laf^2} \sinh \ladh
   \left( \cos \lafh -\frac{\lad^2+\laf^2}{\lad^2-\laf^2} \right) +\frac{\lc}{\laf}\cosh \ladh \sin \lafh \\
 \frac{\lad^2-\laf^2}{2 \lc \lad}\sinh \ladh \left( \cos \lafh+\frac{\lad^2+\laf^2}{\lad^2-\laf^2} \right) - \frac{\laf}{\lc} \cosh \ladh \sin \lafh & \frac{\lad^2-\laf^2}{2 \laf \lad}\sinh \ladh \sin \lafh + \cosh \ladh \cos \lafh \\
\end{array}
\right) \\
& \\
M_{y,\mathrm{F}} & = \left(
\begin{array}{cc}
 -\frac{\lad^2-\laf^2}{2 \laf \lad} \sin \ladh \sinh \lafh + \cos \ladh \cosh \lafh & -\frac{\lc (\lad^2-\laf^2)}{2 \lad \laf^2} \sin \ladh \left( \cosh \lafh - \frac{\lad^2+\laf^2}{\lad^2-\laf^2} \right) + \frac{\lc}{\laf} \cos \ladh \sinh \lafh \\
\frac{\lad}{\lc} \sin \ladh  \left(\frac{\laf^2}{\lad^2} \sinh^2 \laff - \cosh^2 \laff \right ) + \frac{\laf}{\lc} \cos \ladh \sinh \lafh & -\frac{\lad^2-\laf^2}{2 \laf \lad} \sin \ladh \sinh \lafh + \cos \ladh \cosh \lafh \\
\end{array}
\right) \, , 
\end{split}
\end{equation}
where the subscript indicates the type of quadrupole that has been considered as the first element to compute the matrix. Based on these expressions, one obtains the following relationship between the quadrupolar gradients and the phase advances over the cell
\begin{equation}
    \begin{split}
      2 \cos \mu_{\mathrm{c},x} & = 2 \cos \lafh 
      \cosh \ladh + \frac{\lad^2 - \laf^2}{\laf \lad} \sin \lafh \sinh \ladh \\  
      2 \cos \mu_{\mathrm{c},y} & = 2 \cos \ladh 
      \cosh \lafh - \frac{\lad^2 - \laf^2}{\laf \lad} \sin \ladh \sinh \lafh  \, .
    \end{split}
    \label{eq:CFtunes}
\end{equation}
In the results presented in the rest of the paper, these relationships have been used to express the quadrupolar gradients as a function of the cell phase advances, which are the most convenient physical observables to describe the properties of the combined-function periodic structure. This is possible by using Eq.~\eqref{eq:CFtunes} to determine the functions $\laf=f_\mathrm{F}(\mu_{\mathrm{c},x}, \mu_{\mathrm{c},y})$ and $\lad=f_\mathrm{D}(\mu_{\mathrm{c},x}, \mu_{\mathrm{c},y})$, \eg by expressing $f_\mathrm{F}, f_\mathrm{D}$ as truncated power series.

\begin{figure}[htb]
\centering
\includegraphics[width=0.49\textwidth]{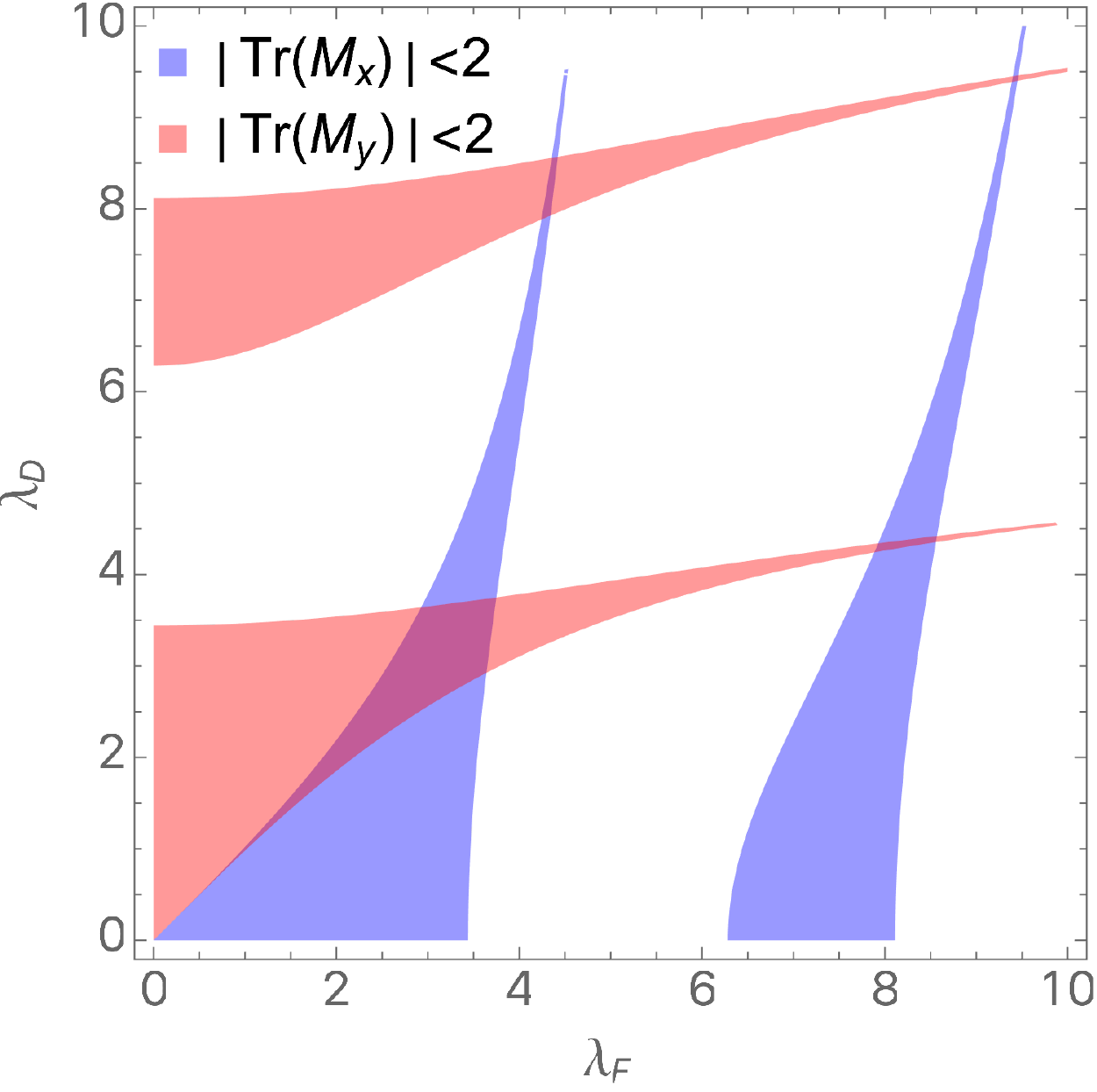}
\includegraphics[width=0.475\textwidth]{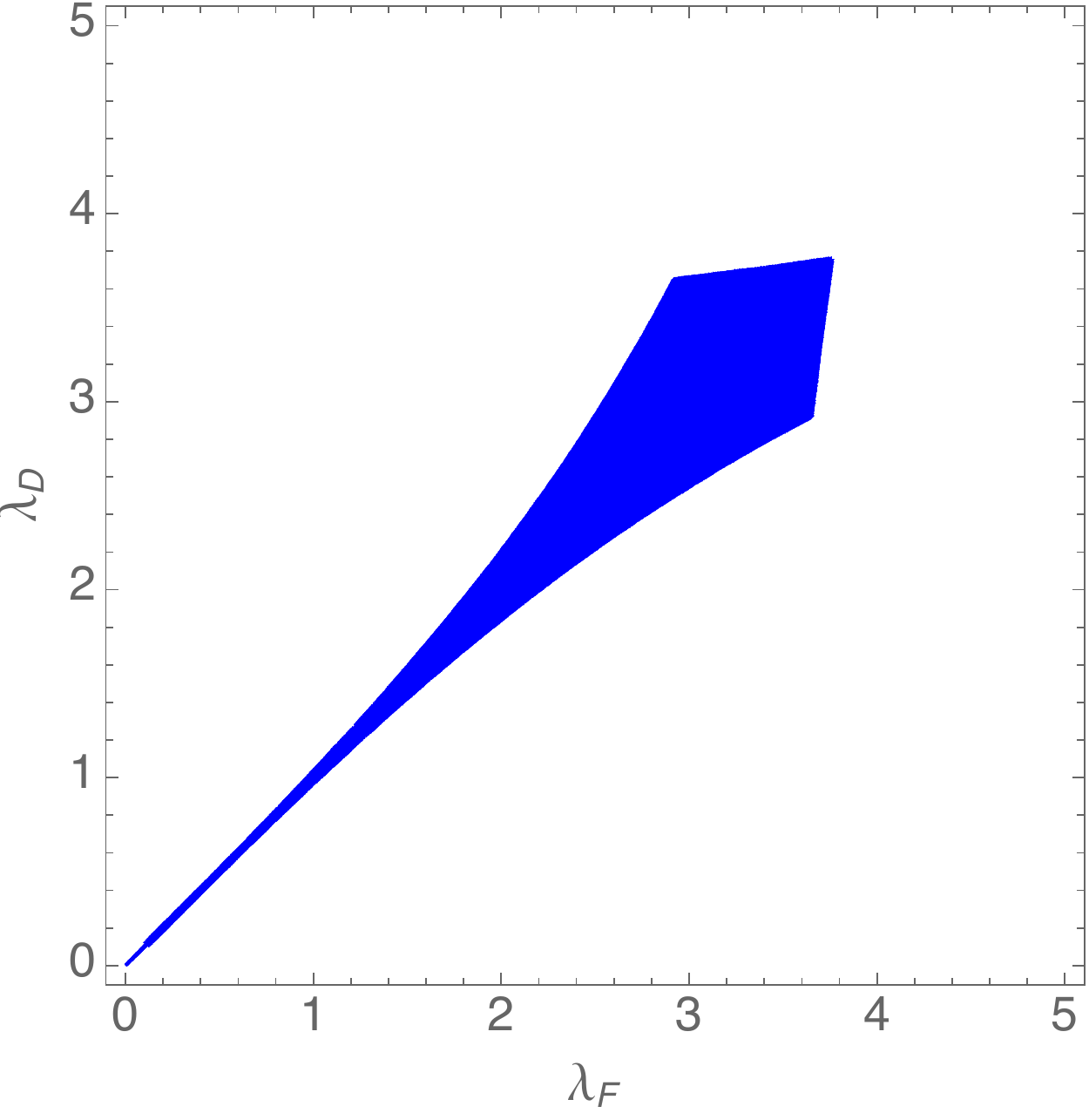}
\caption{Stability region in the $(\laf,\lad)$ space (left) and a zoom-in view of the intersection region (right).}
\label{fig:stability}
\end{figure}

The first point to consider is the stability of the transfer matrix~\cite{Courant:593259}, which corresponds to determining when
\begin{equation}
    | \Tr \left (M_x(\laf,\lad) \right )| \leq 2 \qquad \text{and} \qquad | \Tr \left ( M_y(\laf,\lad) \right )|  \leq 2 \, .
    \label{eq:stabcond}
\end{equation}
Figure~\ref{fig:stability} shows the domain in the space $(\laf,\lad)$ in which the conditions~\eqref{eq:stabcond} are satisfied.
In the left graph, the regions corresponding to the stable conditions for each of the two matrices $M_x, M_y$ independently are depicted, whereas in the right graph, a zoom in is reported, in which the intersection of the two regions is shown. The practical area of stability lies in the domain $0 \leq \laf, \lad \leq 4$ as outside this there are only tiny regions in which the stability is ensured for both horizontal and vertical matrices. Such outlying regions are to be discarded for two main reasons, namely, because the magnetic gradients needed are unnecessarily large, and because the size of the regions is too small to enable stable operation of a particle accelerator. 

It is instructive to derive the relationship between the integrated gradients for a FODO cell (using the thin lens approximation) and those for a combined-function cell, under the simplifying assumption of equal phase advances in both transverse planes. Indeed, one finds by using Eqs.~\eqref{eq:grad_mu} and~\eqref{eq:CFtunes} that
\begin{equation}
    \begin{split}
    \sin \frac{\muc}{2} & = \frac{1}{4} \mathcal{K}_\mathrm{FODO} \, \lc \\
    \cos \muc & = \cos \frac{\sqrt{\mathcal{K}_\mathrm{CF} \, \lc }}{2} \cosh \frac{\sqrt{\mathcal{K}_\mathrm{CF} \, \lc }}{2} 
    \end{split}
\end{equation}
where the symbol $\mathcal{K}$ is used to indicate the integrated normalised gradients~\footnote{The length of the quadrupoles in the FODO and in the combined-function cells are not the same.}. Therefore, one finds
\begin{equation}
    \begin{split}
        \mathcal{K}_\mathrm{FODO} & = \frac{2\sqrt{2}}{\lc} \sqrt{1- \cos \frac{\sqrt{\mathcal{K}_\mathrm{CF} \, \lc }}{2} \cosh \frac{\sqrt{\mathcal{K}_\mathrm{CF} \, \lc }}{2} } \\
        & = \frac{\sqrt 3}{6} \mathcal{K}_\mathrm{CF} \left (1- \frac{1}{13440} \left ( \mathcal{K}_\mathrm{CF} \, \lc  \right )^2 + \mathcal{O} \left ( \mathcal{K}_\mathrm{CF} \, \lc  \right )^{13/2} \right )  \, ,
    \end{split}
\end{equation}
which indicates a nonlinear relationship between the integrated normalised gradients of the FODO and the combined function cell with a sizeable increase of the latter with respect to the former, as all neglected terms are of even power and of negative sign.

The standard parametrisation of the cell matrix by means of the Twiss parameters~\cite{Courant:593259} allows determining the values of the beta-function and dispersion at key locations, namely at the beginning of the cell and at $\lc/2$, which are selected for the sake of comparison with the corresponding Twiss parameters of the FODO cell. Noting that, for symmetry reasons, the alpha-function is zero at these locations, one obtains
\begin{align}
    \beta_{x,\mathrm{F}} & = \frac{(M_{x,\mathrm{F}})_{1,2}}{\sin \mu_{\mathrm{c},x}}  &
    \beta_{y,\mathrm{F}} & = \frac{(M_{y,\mathrm{F}})_{1,2}}{\sin \mu_{\mathrm{c},y}} \\
    \beta_{x,\mathrm{D}} & = \frac{(M_{x,\mathrm{D}})_{1,2}}{\sin \mu_{\mathrm{c},x}}  &
    \beta_{y,\mathrm{D}} & = \frac{(M_{y,\mathrm{D}})_{1,2}}{\sin \mu_{\mathrm{c},y}} \, , 
\end{align}
where $M_{x,\mathrm{D}}$ and $M_{y, \mathbf{D}}$ represent the cell matrices in the horizontal and vertical planes respectively, computed using as starting point of the cell the mid-point of the defocusing quadrupole. Clearly, the trigonometric functions of the cell phase advance can be replaced by the solutions of Eq.~\eqref{eq:CFtunes} in order to provide the beta-functions dependent only on the quadrupole gradients. A first observation is that from the expression of the matrix element one obtains $\beta_{z,\mathrm{F/D}}=\lc \, g(\laf,\lad)=\lc \, \hat{g}(\mu_{\mathrm{c},x},\mu_{\mathrm{c},y})$. 

The dispersion function can be determined using the formalism of $3\times 3$ matrices to deal with the off-momentum dynamics (see \eg \cite{bryant_johnsen_1993}), and it reads
\begin{align}
    D_{x,\mathrm{F}} & = \lc \phc \frac{\lad^2 \sinh \frac{\lad}{4} \cos \frac{\laf}{4}-\laf^2 \sinh \frac{\lad}{4} -\laf \lad \cosh \frac{\lad}{4} \sin \frac{\laf}{4}-\lad^2 \sinh \frac{\lad}{4}}{\lad \laf^2 \rho  \left(\lad \sinh \frac{\lad}{4} \cos \frac{\laf}{4}-\laf \cosh
   \frac{\lad}{4} \sin \frac{\laf}{4} \right)} \\
    D_{x,\mathrm{D}} & = \lc \phc \frac{\laf^2 \cosh \frac{\lad}{4}-\laf \lad \sinh \frac{\lad}{4} \cot \frac{\laf}{4}-\laf^2 -\lad^2}{\lad^2 \laf \rho  \left(\lad \sinh \frac{\lad}{4} \cot \frac{\laf}{4}-\laf \cosh \frac{\lad}{4}\right)} \, .
\end{align}
Also in this case it is worth noting that the dispersion can be expressed as $D_{z,\mathrm{F/D}}=\lc \phc h(\laf,\lad)=\lc \phc \hat{h}(\mu_{\mathrm{c},x},\mu_{\mathrm{c},y})$.

Finally, the chromaticity can be computed by simply applying its definition to Eq.~\eqref{eq:CFtunes}, and one obtains
\begin{equation}
    \begin{split}
  \mu'_{\mathrm{c},x} & = -\frac{1}{8 \sin \mu_{\mathrm{c},x}} \left [\frac{\left(\laf^2-3 \lad^2\right)}{\lad} \sinh \frac{\lad}{2} \cos \frac{\laf}{2} +\frac{\left(3 \laf^2-\lad^2\right)}{\laf} \cosh \frac{\lad}{2} \sin \frac{\laf}{2} \right ] \\
   \mu'_{\mathrm{c},y} & = -\frac{1}{8 \sin \mu_{\mathrm{c},y}} \left [ \frac{\left(\lad^2-3 \laf^2\right)}{\laf} \cos \frac{\lad}{2} \sinh \frac{\laf}{2} + \frac{\left(3 \lad^2-\laf^2\right)}{\lad} \sin \frac{\lad}{2} \cosh \frac{\laf}{2} \right ] \, .   
    \end{split}
\end{equation}

A discussion of the optical parameters that have an impact on the longitudinal beam dynamics is reported in Appendix~\ref{sec:app2}
\subsection{Comparison of optical parameters for FODO and combined-function cells} \label{sec:optcomparison}
\begin{figure}[htb]
\centering
\includegraphics[width=0.59\textwidth]{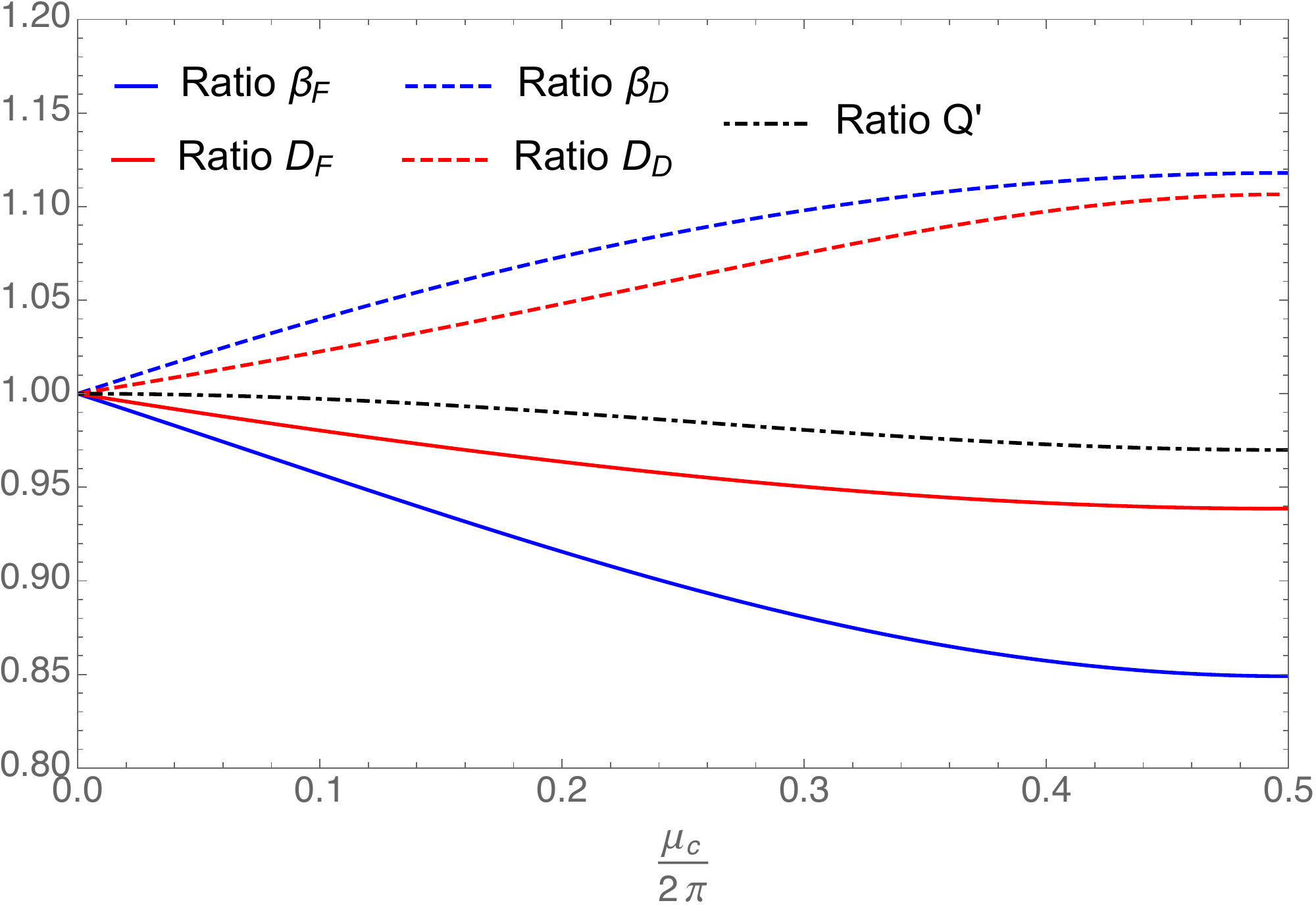}
\caption{Ratio of the optical parameters, dispersion, and chromaticity of the combined-function periodic cell with respect to the corresponding parameters of the FODO cell as a function of the cell phase advance.}
\label{fig:comparison}
\end{figure}

A better understanding of the properties of the combined-function periodic cell can be obtained by carrying out a direct comparison of the optical parameters and of the dispersion function. To this aim, a simplification has been introduced by setting $\laf=\lad$, which is not a serious limitation for the analysis performed, given that in most cases the horizontal and vertical phases advances of the FODO cells of modern colliders are indeed the same. The comparison is made by looking at the cell parameters as a function of the cell phase advance and the results are summarised in Fig.~\ref{fig:comparison}, where the ratios of the parameters for the combined function periodic cell to the corresponding parameters of the FODO cell are shown as a function of the cell phase advance.
The ratios of the key optical parameters of the proposed cell show some nice features. Both $\beta_\mathrm{F}$ and $D_\mathrm{F}$ are smaller than the corresponding values for the FODO cell. When $\muc/(2\pi)=1/4$, \ie for a $90^\circ$ degrees phase advance, a very common value in accelerator design, $\beta_\mathrm{F}$ is about $10$\% smaller, whereas $D_\mathrm{F}$ is about $5$\% smaller than for the case of a FODO cell. Although this reduction is not spectacular, it goes in the right direction, reducing the needs in terms of beam aperture. At the same time, $\beta_\mathrm{D}$ and $D_\mathrm{D}$ are a bit larger than in the case of a FODO cell. Globally, this means that the beta function varies less than in a FODO cell and this has an impact on the chromatic properties of the proposed combined-function cell. This observation is connected with the ratio of the chromaticities, which is slightly below unity over the whole range of phase advances considered in this study. These properties have  implications for the chromaticity correction and deserve a detailed  discussion that is presented in the next section.
\subsection{Chromaticity correction} \label{sec:chromcorr}
\begin{figure}[htb]
\centering
\includegraphics[width=0.59\textwidth]{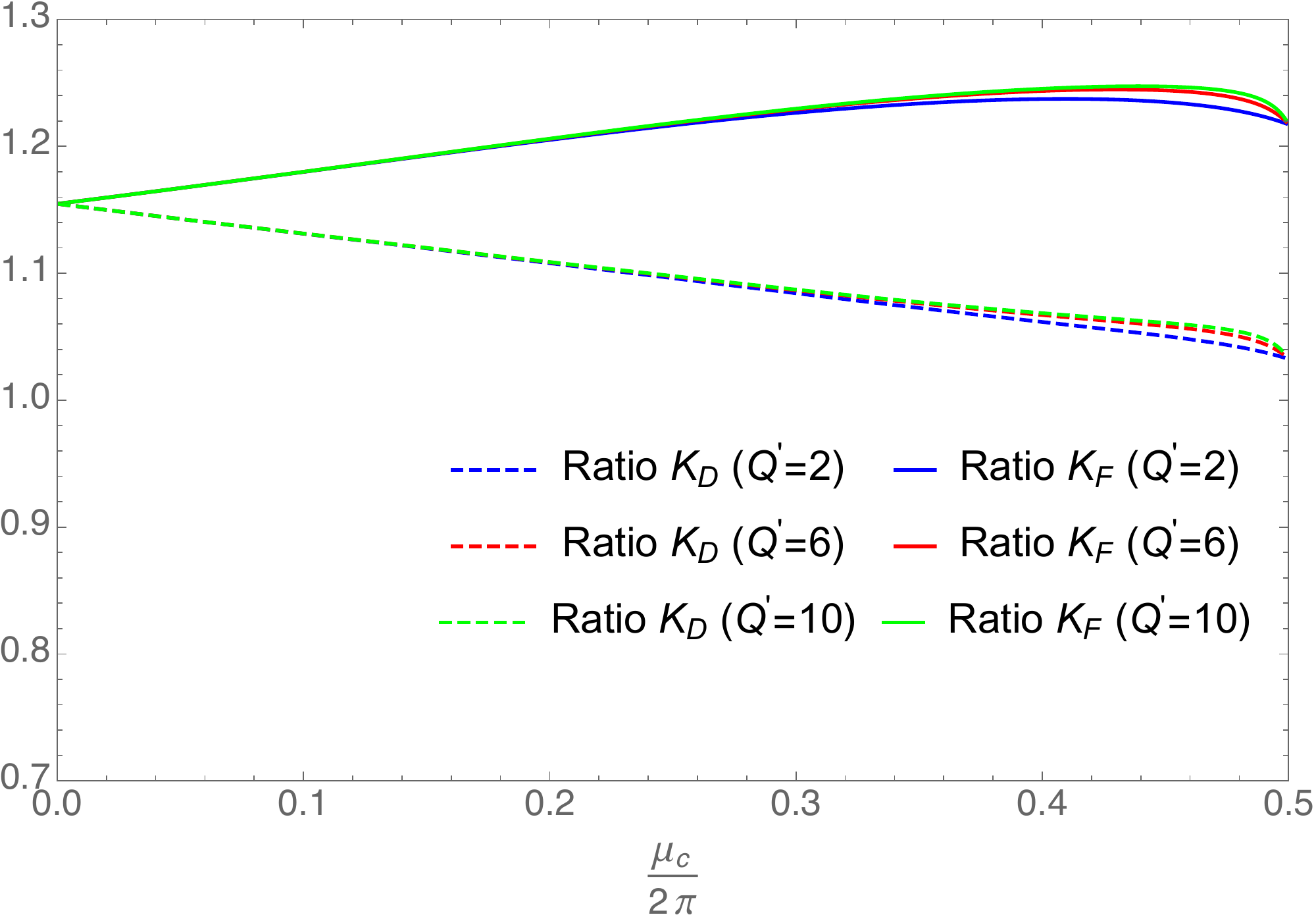}
\caption{Ratio of the strength of the sextupoles used for the correction of the chromaticity for the combined-function periodic cell to the FODO cell as a function of the cell phase advance.}
\label{fig:comp_chrom}
\end{figure}

In Appendix~\ref{sec:app1}, the topic of chromatic corrections for a FODO cell is considered and the main formulas are derived in detail (see Eqs.~\eqref{eq:chromcorr0}--\eqref{eq:chrom_corr3}). A similar approach can also be applied to the combined-function cell, by assuming that two sextupoles are installed at the beginning and midpoint of the cell, respectively. A direct comparison of the strength of the two sextupoles is shown in Fig.~\ref{fig:comp_chrom}, where the ratio of the sextupole strength for the combined-function cell to that of the FODO cell is shown for three values of the target chromaticity and in dependence of the cell phase advance. 
The sextupole strength required for chromaticity correction in the combined-function cell is different from the standard FODO cell, showing decreasing or increasing trends for defocusing or focusing sextupoles respectively. This is not a major issue, as around the $90^\circ$ degrees phase advance value, the extra strength only reaches a value of about $20$\%, for the focusing sextupoles, and about $10$\% for the defocusing sextupoles. Note that the large value of $Q'$ used in the simulations is justified by the LHC experience~\cite{Metral:2199121}, in which the machine is operated with $15 \leq Q' \leq 20$ between the injection and collision stages, to stabilise the beam. Hence, this value is considered typical for future hadron colliders~\cite{FCC-hhCDR}. 
Alternative scenarios have been considered, which are based on an asymmetric combined-function cell, to assess whether a lower strength of the sextupole magnets can be obtained. The asymmetric combined-function cell layout has been considered in detail in Appendix~\ref{sec:app3} and consists of sections of focusing and defocusing magnets with different total lengths. This introduces an additional parameter that can be used to control the optical properties of the cell. Of course, this implies that the combined-function magnets are of two types featuring different nominal quadrupolar fields. 
\begin{figure}[htb]
\centering
\includegraphics[height=0.30\textwidth,clip=]{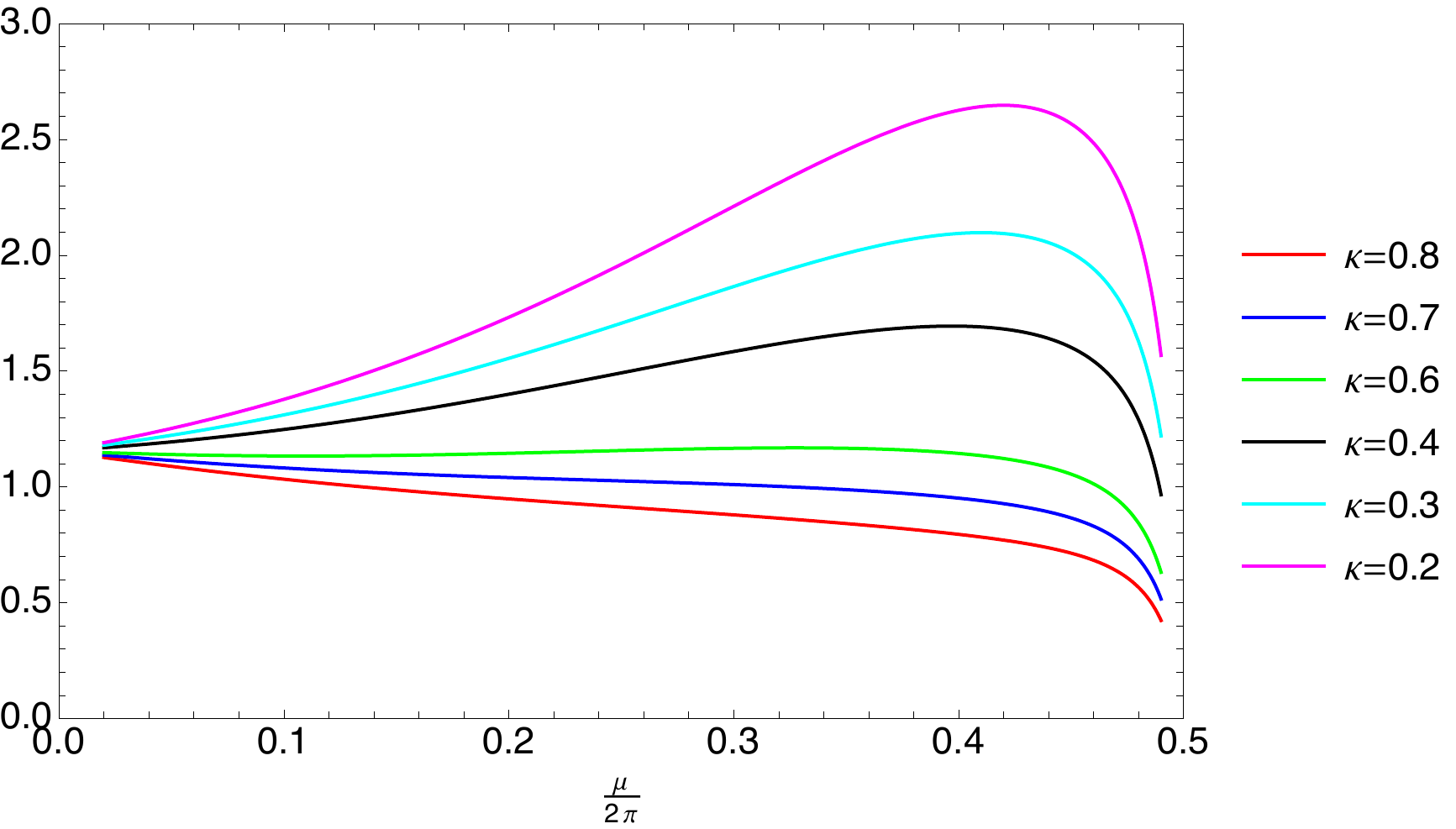}
\includegraphics[trim=0mm 0mm 30mm 0mm,height=0.30\textwidth,clip=]{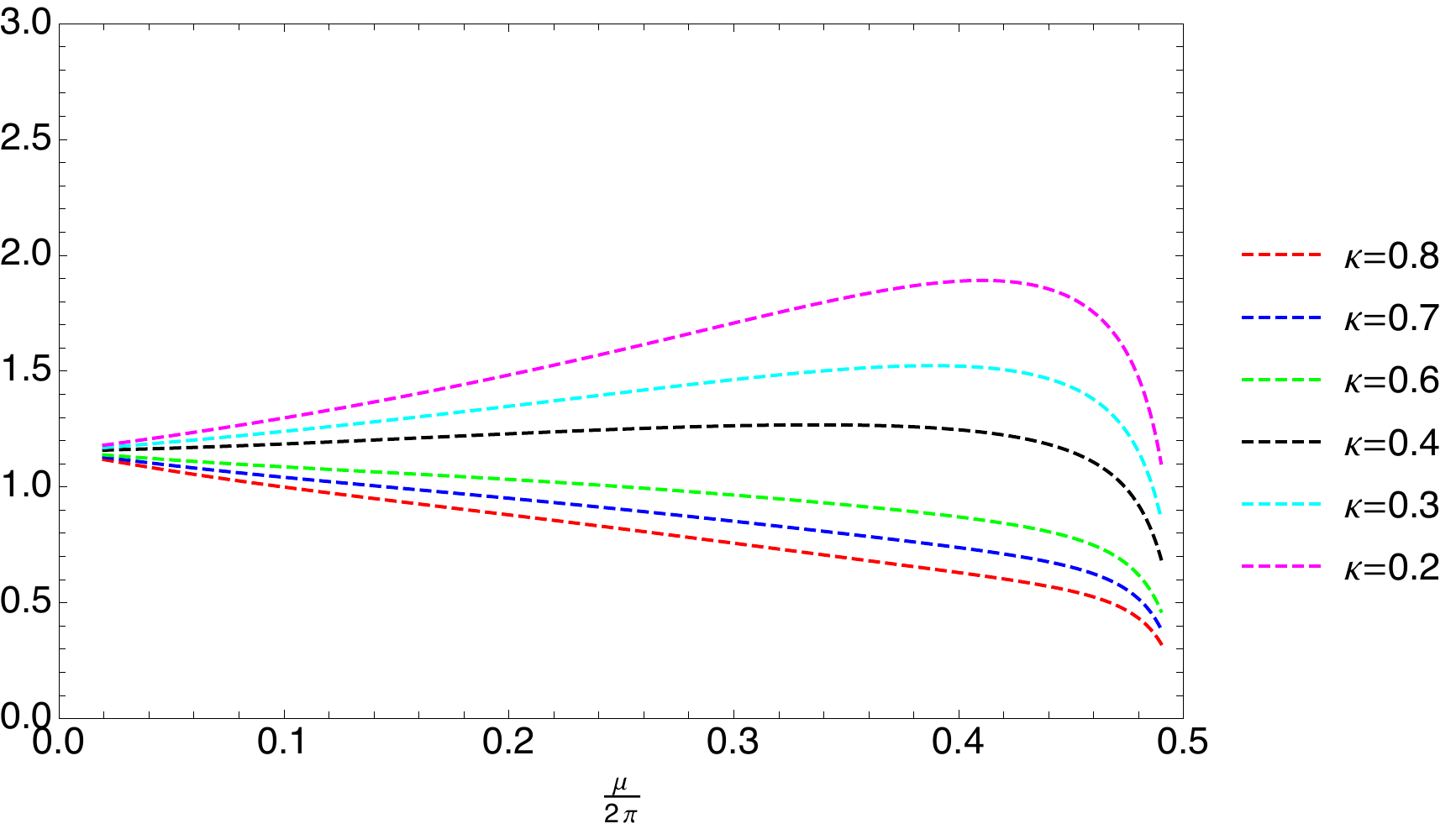}
\includegraphics[height=0.30\textwidth,clip=]{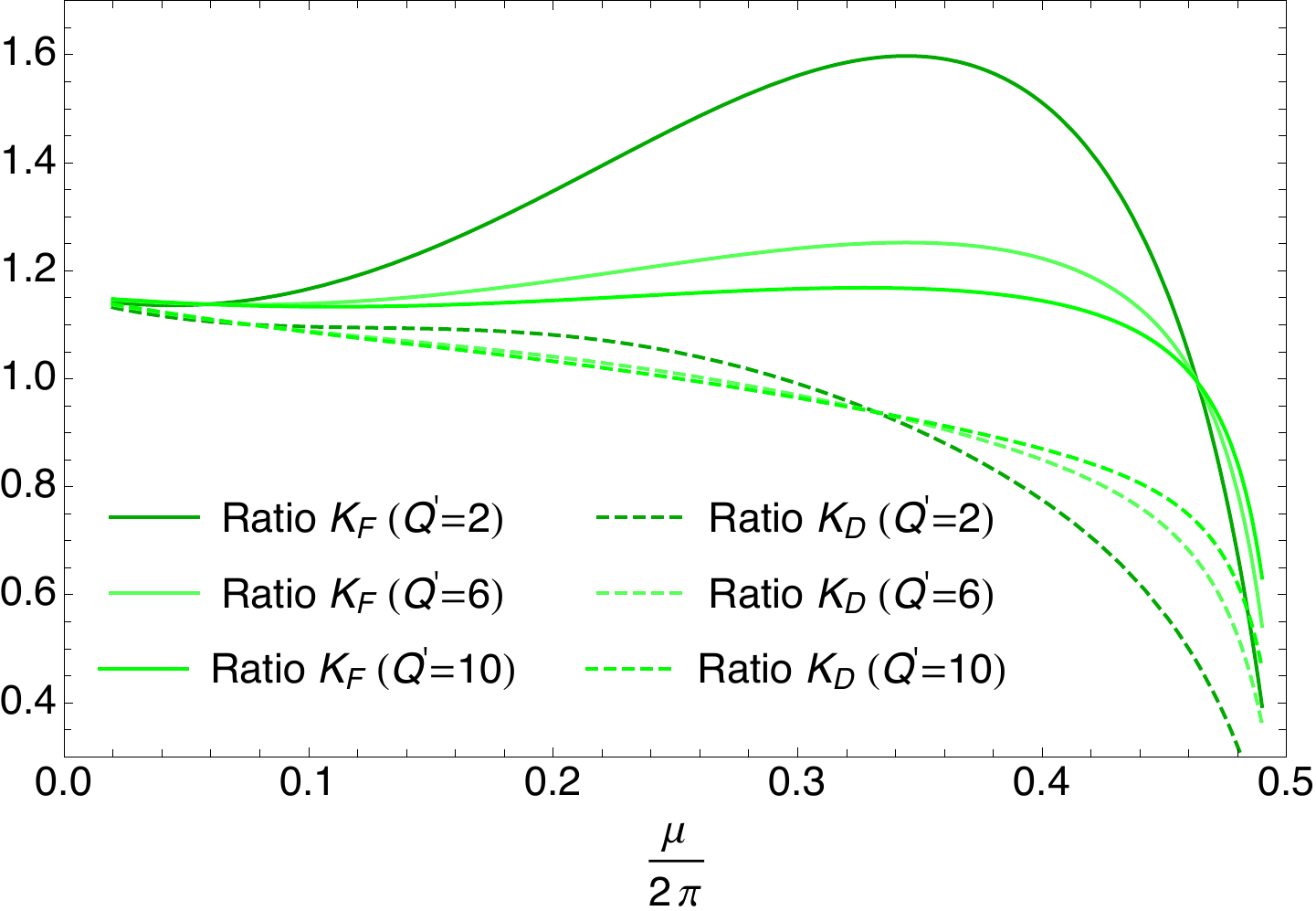}
\caption{Upper: ratio of the strength of the focusing (left) and defocusing (right) sextupoles for an asymmetric combined-function to the corresponding values for a FODO cell layout as a function of the phase advance. For this particular case, the target chromaticity value is set to $10$ and the phase advance is the same for the two transverse planes. Lower: ratio of strength of the focusing and defocusing sextupoles for an asymmetric cell layout to the corresponding values for a FODO cell layout with $\kappa=0.6$ and for different values of the target chromaticity all as a function of the phase advance.}
\label{fig:CFchromcorr}
\end{figure}

The results of the simulations of the chromaticity corrections are summarised in Fig.~\ref{fig:CFchromcorr}, in which the ratios of the quantities for asymmetric combined-function cell layouts to the same quantities for a FODO cell layout are depicted. In the upper plots, the strength of the focusing (left) and defocusing (right) sextupoles are shown as a function of the cell phase advance, which is set equal in the horizontal and vertical planes. Several values of the parameter $\kappa$, characterising the asymmetry of the cell layout, have been considered and it is clear that the solutions for $\kappa > 0.5$ are the most interesting ones as they feature a reduced need of sextupoles strength. In the lower plot, the case of $\kappa=0.6$, which is the most promising, is shown. The ratio of strength of the focusing and defocusing sextupoles are shown for three different values of the target chromaticity and over the whole interval of the cell's phase advance. The defocusing sextupoles feature a strength that is rather close, less than $1$\% stronger, to that of the FODO cell for the phase advance close to $90^\circ$ degrees. The focusing sextupoles are about $20$\% stronger than the corresponding elements of the FODO cell, but feature a peculiar behaviour as a function of the target chromaticity, namely, a rather important increase in strength is observed when correcting towards a low value of chromaticity. This is linked with the asymmetric layout that generates a preferred interval of chromaticity values. This is not a serious issue in itself, given that for beam stability considerations, these values are not needed in routine operation. Still, this observation means that the proposed alternative solution, although effective in reducing the strength of the chromatic sextupoles with respect to the symmetric combined-function layout, might limit the optics flexibility of the periodic cell.

\begin{figure}[htb]
\centering
\includegraphics[width=0.47\textwidth,clip=]{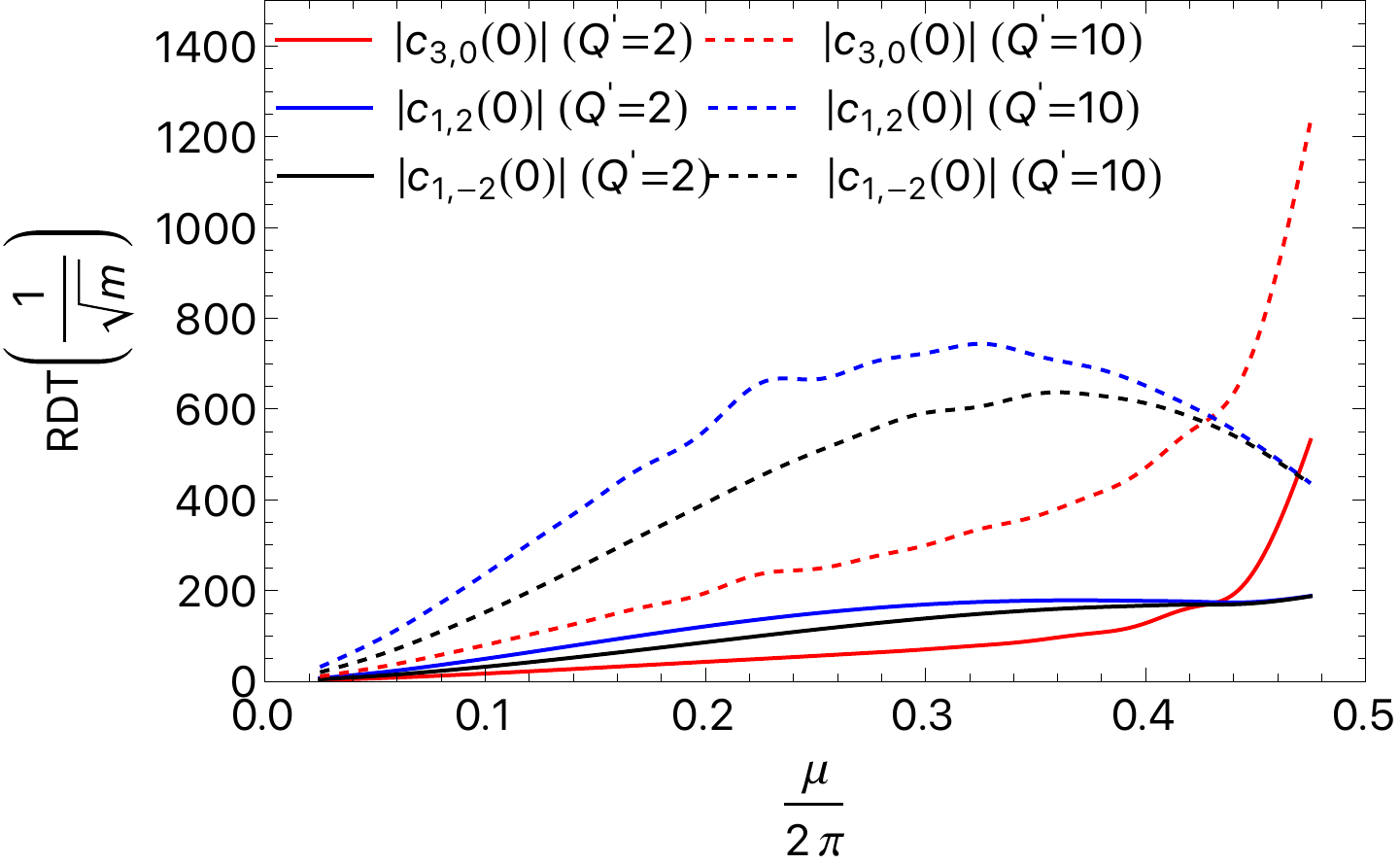}
\includegraphics[width=0.47\textwidth,clip=]{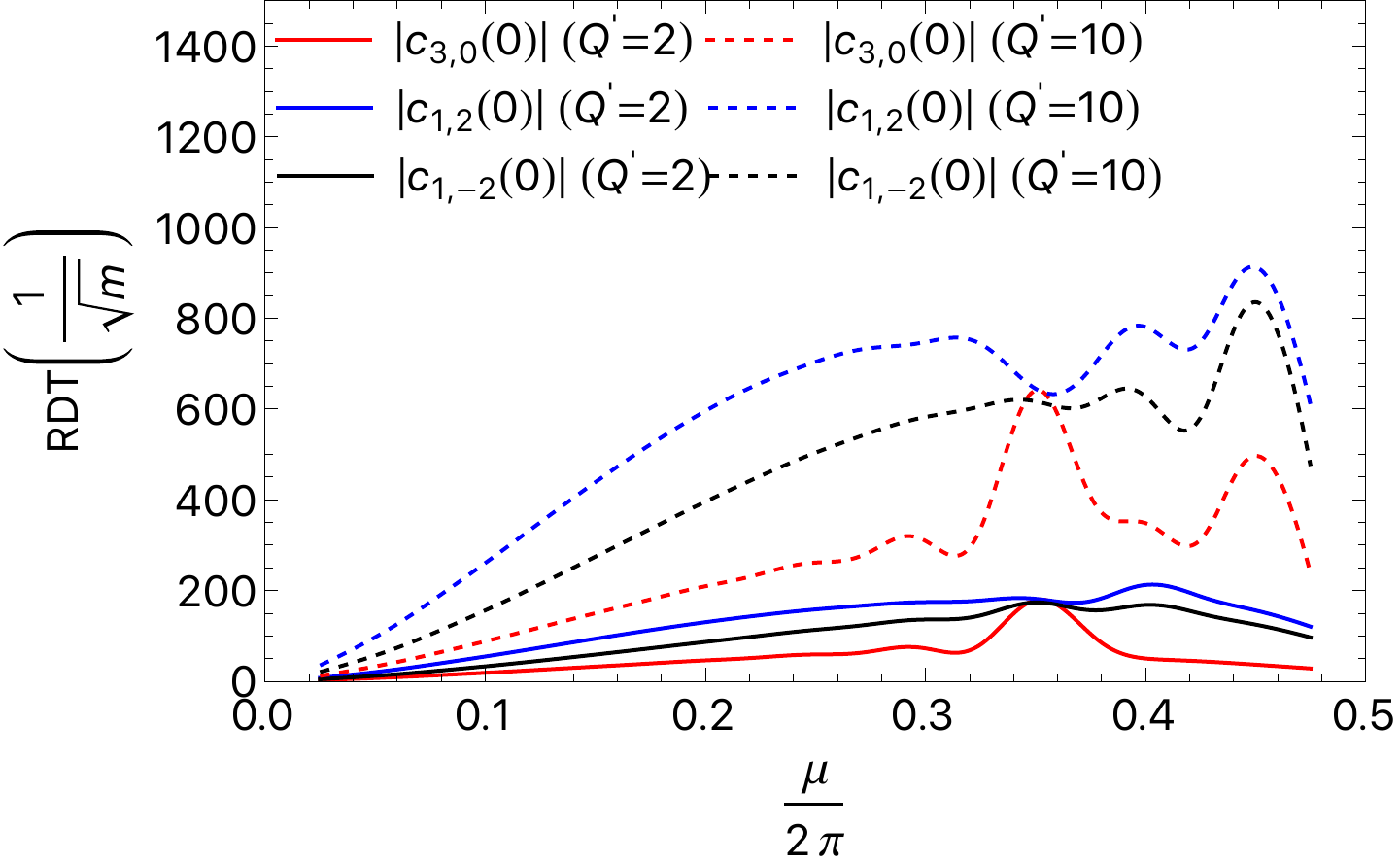}
\includegraphics[width=0.47\textwidth,clip=]{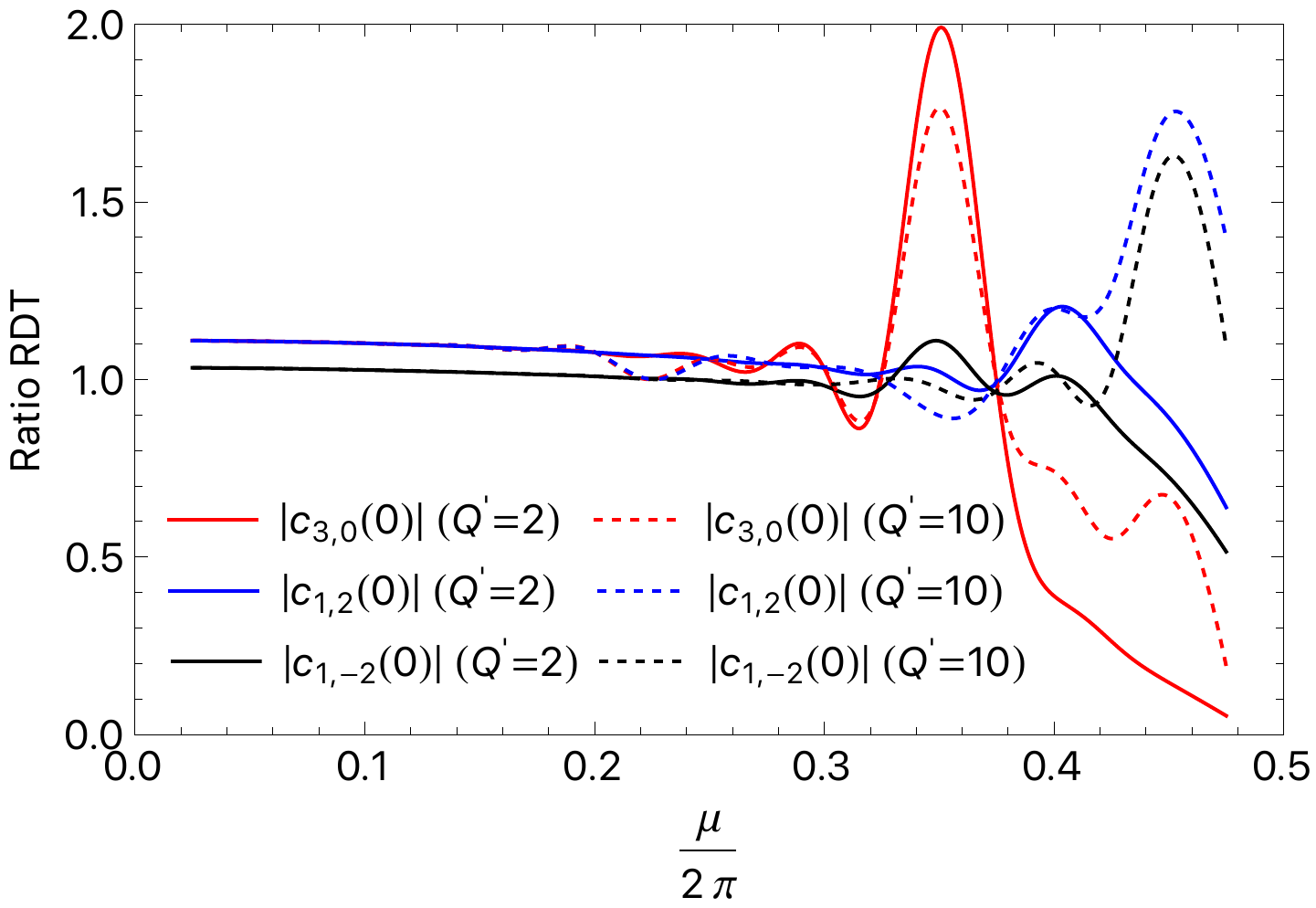}
\caption{Upper: sextupolar RDTs for the standard FODO cell (left) and the symmetric combined-function cell (right) as a function of the cell phase advance and for two values of the chromaticity. Lower: ratio of the sextupolar RDTs of the combined-function to FODO cells. The RDTs have very similar values for the value of the cell phase advance up to $0.3$, which covers most of the cases relevant for applications.}
\label{fig:compRDT}
\end{figure}

As an additional verification, and to compare the performance of the FODO and symmetric combined-function cell designs, the RDTs have been evaluated and are shown in Fig.~\ref{fig:compRDT} for the FODO (top-left) and symmetric combined function (top-right) cells, along with their ratio (bottom). The comparison was carried out using MAD-X~\cite{madx} and PTC~\cite{Schmidt:573082} to evaluate the RDTs using sample cells of $200$~m length, one implementing a thin-lens version of a standard FODO cell and one the proposed combined-function cell. 

Both FODO and combined-function cell configurations feature RDTs that are very similar for values of the cell phase advance $\muc/2\pi < 0.3$, which covers the interesting values for applications. This consideration holds for both chromaticity values tested with the numerical simulations. Therefore, we conclude that there are no sizeable differences in terms of key optical properties between the standard FODO and the proposed combined-function cells.
\section{A possible combined-function cell for a high-energy collider} \label{sec:layout}
\begin{figure}[htb]
\centering
\includegraphics[trim= 3mm 30mm 13mm 22mm, angle=-90,width=0.49\textwidth,clip=]{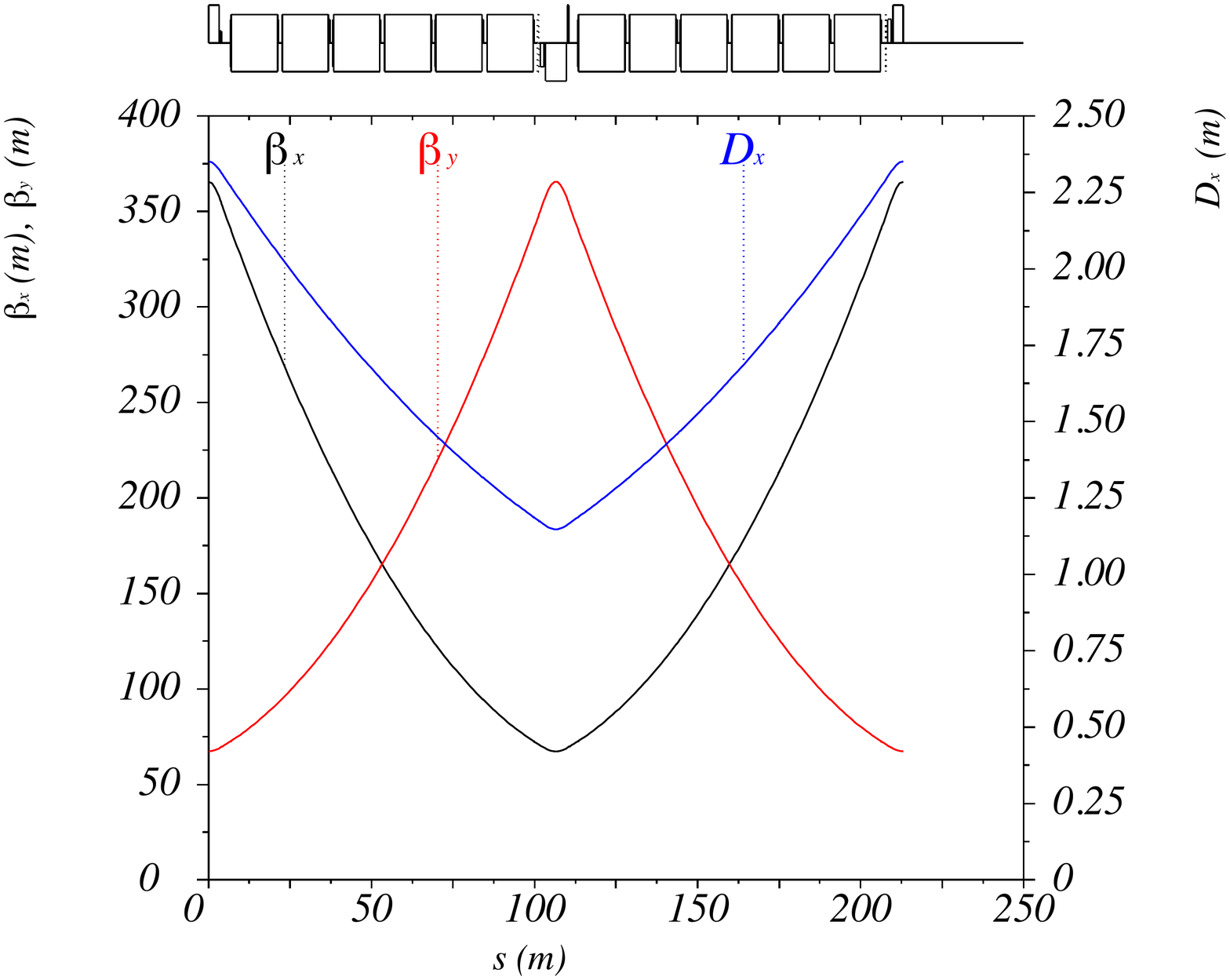}
\includegraphics[trim= 3mm 30mm 13mm 22mm,angle=-90,width=0.475\textwidth,clip=]{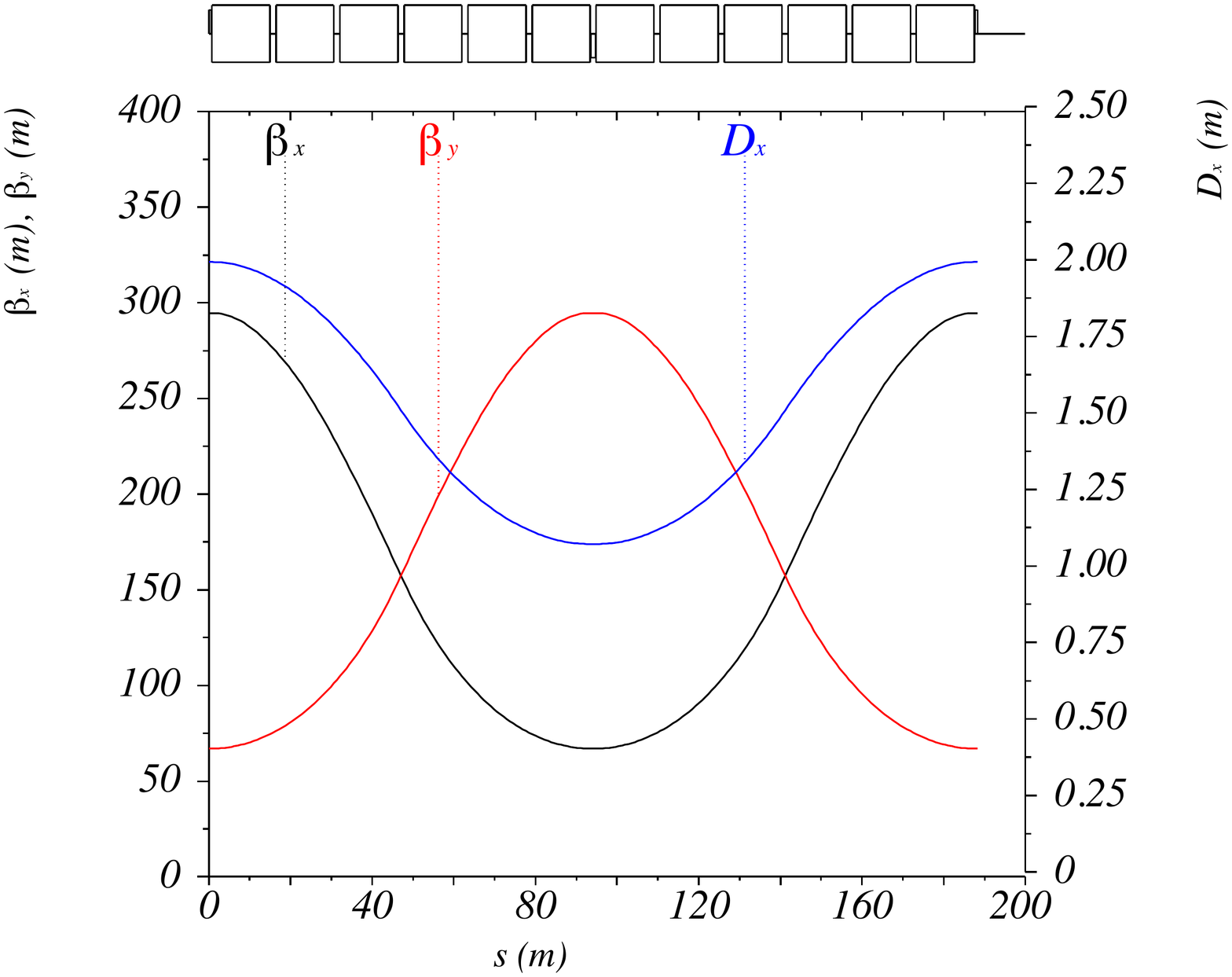}
\caption{Optical parameters of the standard FODO cell of the FCC-hh ring  (left) and of the proposed combined-function cell (right).}
\label{fig:realistic_cells}
\end{figure}

The concepts developed in the previous sections can be used to propose a periodic-cell design for a high-energy circular collider. To this aim, we will use the case of the FCC-hh ring. In Fig.~\ref{fig:realistic_cells} (left), the optical parameters of the nominal FCC-hh cell, according to the layout described in~\cite{FCC-hhCDR} are shown. Above the plot, the cell layout is visible: the boxes centred around the horizontal line represent the twelve main dipoles,  each of $14.187$~m of magnetic length, whereas the boxes placed above or below the horizontal line represent the two main quadrupoles, each of $6.4$~m of magnetic length. All ancillary devices, such as beam position mornitors and corrector magnets are also included, but not used in these studies. 
For this realistic design, including the length of the interconnections, one obtains $\chi_\mathrm{FODO} = 0.80$ with the non-normalised and non-integrated strength of the main quadrupoles $\hat{K}_{2, \mathrm{F/D}}$ and of the chromatic sextupoles $\hat{K}_{3,\mathrm{F/D}}$ given by
\begin{align}
         \hat{K}_{2,\mathrm{F}}^\mathrm{FODO} & = 353.6\, \text{T}/\text{m} & \hat{K}_{3,\mathrm{F}}^\mathrm{FODO} & = 27848.3\, \text{T}/\text{m}^{2} \\
         \left \vert \hat{K}_{2,\mathrm{D}}^\mathrm{FODO} \right \vert & = 353.6\, \text{T}/\text{m} & \left \vert \hat{K}_{3,\mathrm{D}}^\mathrm{FODO} \right \vert & = 55823.9\, \text{T}/\text{m}^{2} \, .
\end{align}
Note that the absolute strength of the main quadrupoles has been computed to set $\muc=\pi/2$, whereas the absolute strength of the chromatic sextupoles has been determined by correcting the chromaticity to $Q'=10$. Furthermore, all values quoted here are non-normalised and refer to operation at $50$~TeV, \ie the nominal single beam energy for the $100$~TeV collision energy of the FCC-hh.

In the right plot of Fig.~\ref{fig:realistic_cells}, the optical parameters of the proposed combined-function cell are shown. The layout has been designed using twelve dipoles with the same magnetic properties as those of the FCC-hh cell, with a superimposed quadrupolar component making a sequence of three focusing, six defocusing, and three focusing combined-function magnets. The distance between the combined function dipoles is the same as that of the nominal FCC-hh cell, \ie $1.5$~m, for the sake of comparison. The chromatic sextupoles are located at the extremities and in the middle of the cell. Under these assumptions, the combined-function cell is shorter than the standard cell, because of the removal of the main quadrupoles and the ancillary magnets that would be located in the same cryostat, and one obtains $\chi_\mathrm{CF}=0.90$, which corresponds to an improvement of almost $13$\%. The observations made about the features of the optical parameters for a combined function cell are fully confirmed, \ie smaller beta- and dispersion functions than for a FODO cell are achieved. Of course, this is due partly to the reduced cell length and partly to the special optical design. In terms of strength of the quadrupolar components in the dipoles and the strength of the chromatic sextupoles, one obtains
\begin{align}
         \hat{K}_{2,\mathrm{F}}^\mathrm{CF} & = 51.2\, \text{T}/\text{m} & \hat{K}_{3,\mathrm{F}}^\mathrm{CF} & = 40263.5\, \text{T}/\text{m}^{2} \\
         \left \vert \hat{K}_{2,\mathrm{D}}^\mathrm{CF} \right \vert & = 51.2\, \text{T}/\text{m} & \left \vert \hat{K}_{3,\mathrm{D}}^\mathrm{CF} \right \vert & = 75831.4\, \text{T}/\text{m}^{2} \, .
\end{align}
It is worth noting that while the direct comparison of the absolute sextupolar strengths for the FODO and combined-function cells reveals an increase of about $40$\% for the latter, whenever the comparison is carried out with a FODO cell of the same length as that of the combined-function cell, then the increase is much more modest, in the order of $20$\% and $10$\% for the focusing and defocusing sextupoles respectively.

The required quadrupolar strength is expressed as a derivative of the field in a form that is more suitable for normal-conducting magnets. Indeed, in the case of superconducting magnets, it is customary to express the magnetic field in terms of multipolar components according to~\cite{LHCDR} 
\begin{equation}
    B_y+iB_x = B_\mathrm{ref} \sum_{n=1}^\infty \left ( b_n + i a_n\right ) \left ( \frac{x+i y}{R_\mathrm{ref}}\right )^{n-1} \, ,
\end{equation}
where $B_\mathrm{ref}$ is the reference field of the magnet for which the multipolar expansion is computed (for the FCC-hh it is $16$~T); $b_n$ and $a_n$ are the so-called normal and skew components, respectively; $R_\mathrm{ref}$ is the reference radius at which the multipolar components are given (for the FCC-hh and the LHC it is $17$~mm); and $n=1$ represents dipolar components, $n=2$ quadrupolar components, and so on. It is easy to derive that
\begin{equation}
    b_2 = \frac{\hat{K}_2}{B_\mathrm{ref}} R_\mathrm{ref}
\end{equation}
and using the LHC convention, \ie with $b_n$ and $a_n$ expressed in units of $10^{-4}$, then $b_{2,\mathrm{CF}} = 544$.

It is worth stressing that at this point of the analysis no further optimisation of the cell layout has been carried out. However, it is clear that increasing the cell length would allow reducing $b_{2,\mathrm{CF}}$, as well as the strength of the chromatic sextupoles, although this would be achieved with a corresponding increase of the values of the optical functions.

\begin{figure}[htb]
\centering
\includegraphics[trim= 3mm 30mm 13mm 22mm, angle=-90,width=0.49\textwidth,clip=]{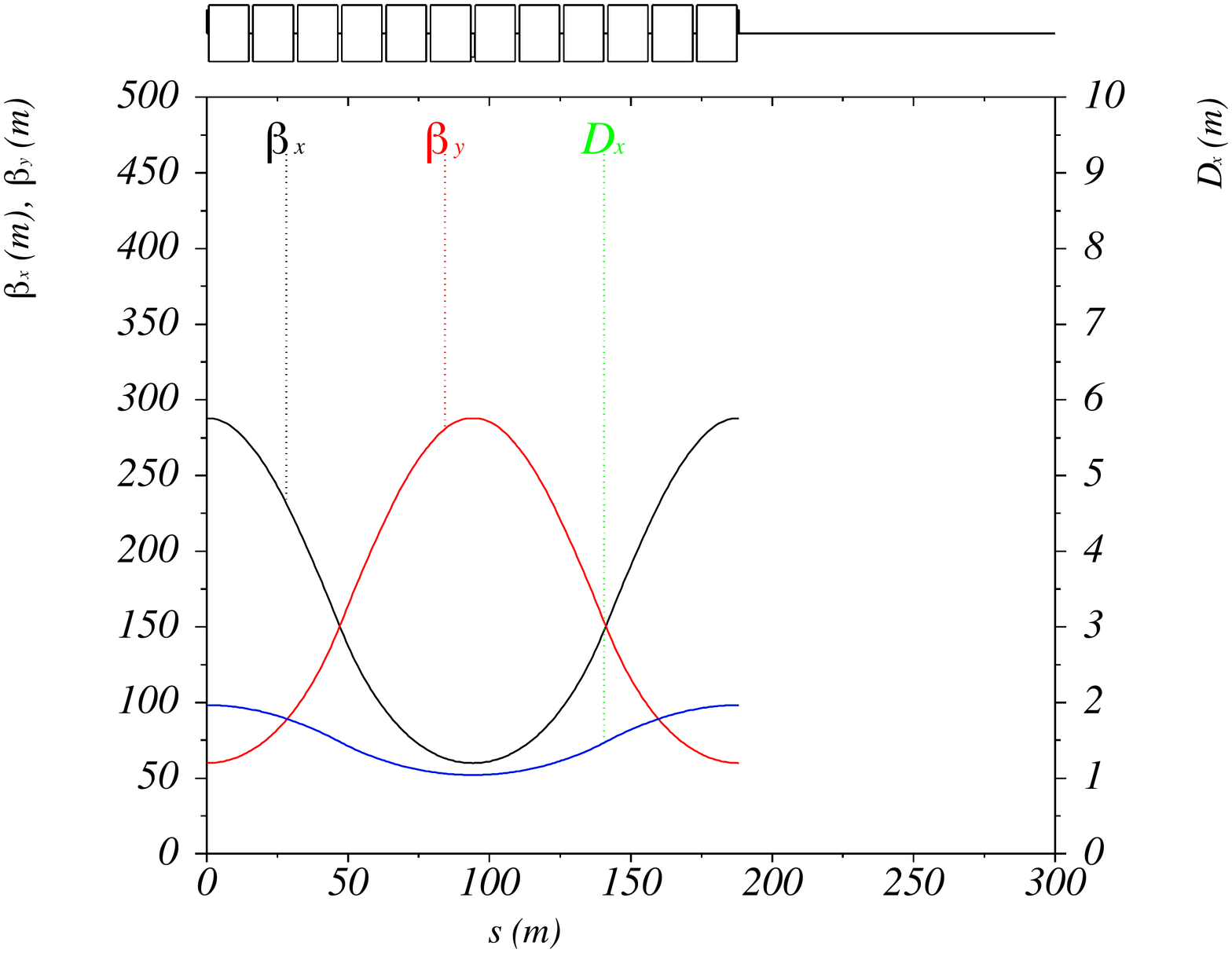}
\includegraphics[trim= 3mm 30mm 13mm 22mm,angle=-90,width=0.475\textwidth,clip=]{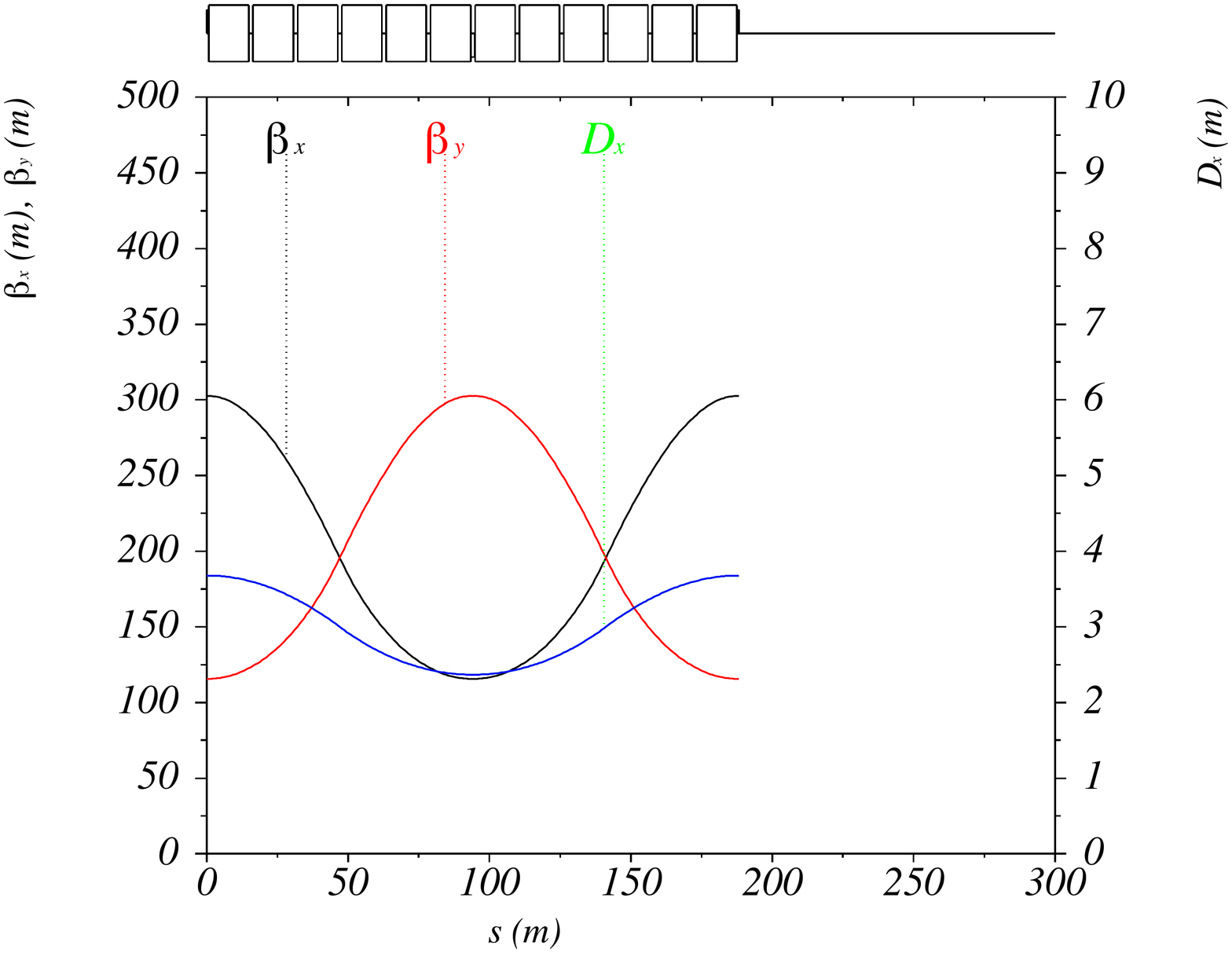}
\includegraphics[trim= 3mm 30mm 13mm 22mm, angle=-90,width=0.49\textwidth,clip=]{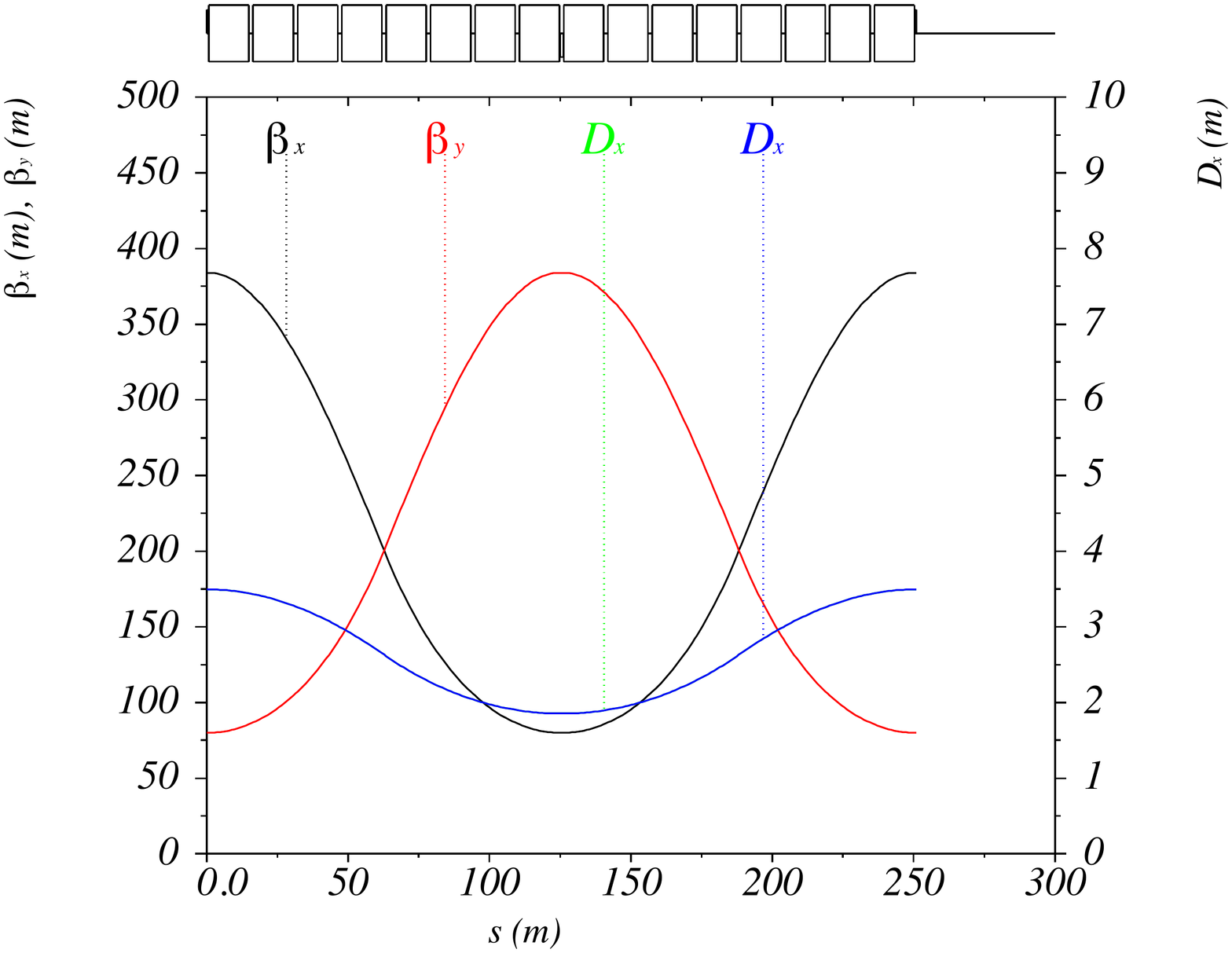}
\includegraphics[trim= 3mm 30mm 13mm 22mm,angle=-90,width=0.475\textwidth,clip=]{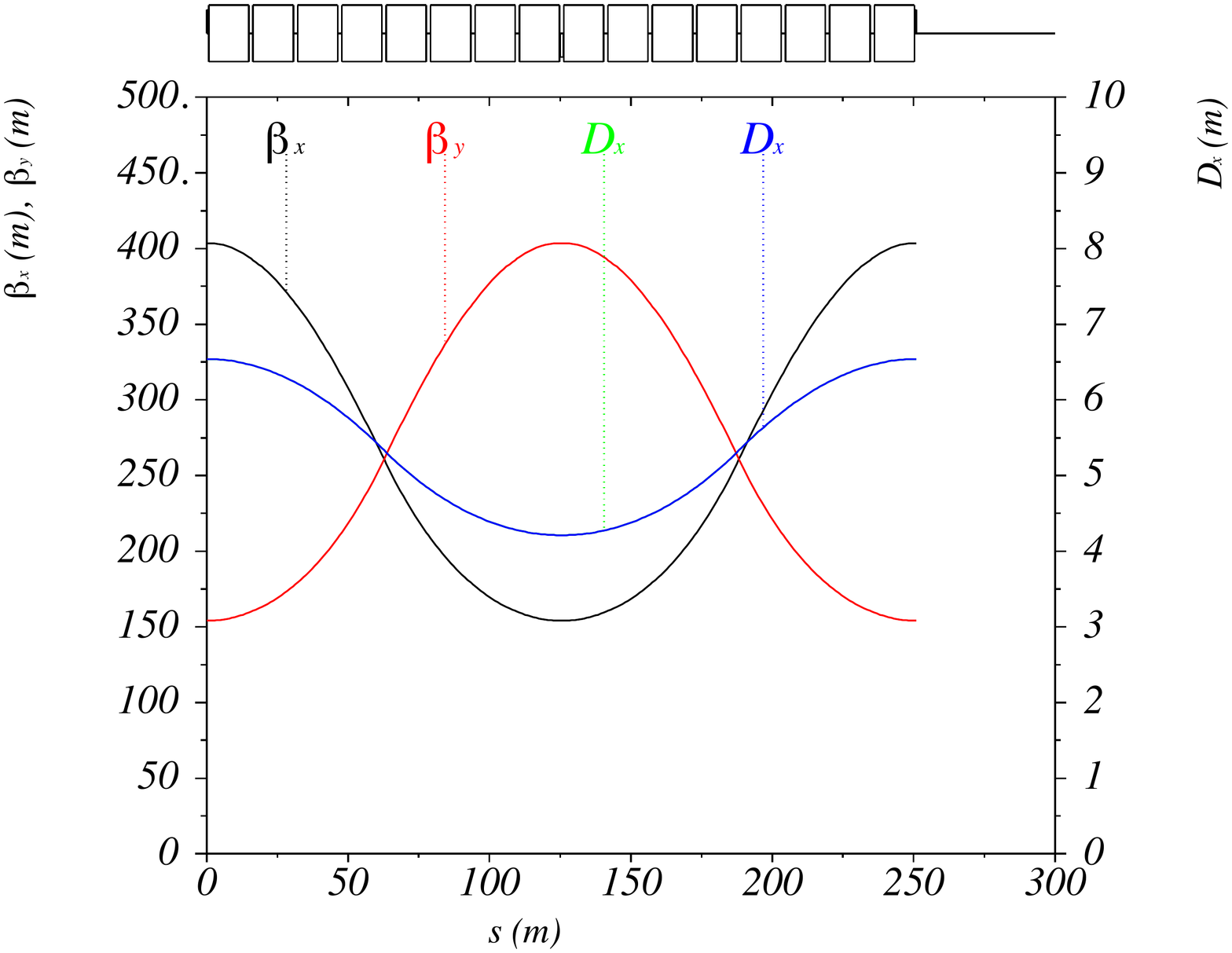}
\caption{Optical parameters of the 12-dipole (top) and 16-dipole (bottom) variants of the combined-function cell. Layouts with $90^\circ$ degrees (left) and $60^\circ$ degrees phase advance (right) are also shown. Note the upper-left plot corresponds to the right one in Fig.~\ref{fig:realistic_cells}.}
\label{fig:variants}
\end{figure}

The combined function layout considered so far is the one that can be directly compared with the nominal FCC-hh cell, as the number of dipoles and the phase advance is the same. However, one could consider an increased number of dipoles, or a different phase advance. Examples of alternative layouts are shown in Fig.~\ref{fig:variants}, where the optical parameters for combined function cells with twelve (top row) or 16 (bottom row) dipoles are shown for a phase advance of $90^\circ$ degrees (left column) and $60^\circ$ degrees (right column).

The increase of beta function and dispersion due to the increased cell length is clearly visible, but rather modest, and could still be acceptable in terms of beam aperture. The change of phase advance also has an impact on the beta-functions as the layout with $60^\circ$ degrees phase advance is further away from the value that generates the minimum beta-functions. However, the impact on the dispersion function is much more relevant, indicating that chromatic aberrations would be more important. This value of phase advance has been tested as it requires a lower quadrupolar gradient and behaves better in terms of dynamic aperture~\cite{PhysRevAccelBeams.23.101602,Nosochkov:IPAC18-MOPMF067}. The main features of the four variants of combined-function cell layouts considered are summarised in Table~\ref{table:summary}.

\begin{table}[htb]
\centering
\caption{Summary of the main features of the four variants of the combined-function periodic cell. All variants feature $\chi=0.90$. The target value of the chromaticity is $Q'=10$.}
\label{table:summary}
\begin{tabular}{cccccc|ccccc}
\hline 
& \multicolumn{10}{c}{Number of dipoles} \\
& \multicolumn{5}{c|}{12 ($\lc= 188.246$~m)} & \multicolumn{5}{c }{16 ($\lc= 250.995$~m)} \\
$\muc$ & $\bma$ & $\dma$ & $\gamma_\mathrm{tr}$ & $\hat{K}_{2}$ & $\hat{K}_{3,\mathrm{F}}/|\hat{K}_{3,\mathrm{D}}|$ & $\bma$ & $\dma$ & $\gamma_\mathrm{tr}$ & $\hat{K}_{2}$ & $\hat{K}_{3,\mathrm{F}}/|\hat{K}_{3,\mathrm{D}}|$ \\
degrees & (m) & (m) & & $\text{T}/\text{m}$ & $\text{T}/\text{m}^{2}$ & (m) & (m) & & $\text{T}/\text{m}$ & $\text{T}/\text{m}^{2}$ \\
\hline
90 & 287.71 & 1.96 & 88.39 & 51.2 & 40263.5/75831.4 & 383.92 & 3.49 & 66.29 & 28.8 & 16960.2/31965.4 \\
60 & 302.49 & 3.68 & 62.23 & 36.1 & 25889.0/40177.5 & 403.53 & 6.54 & 46.67 & 20.3 & 10907.1/16935.0\\
\hline \hline
\end{tabular}
\end{table}

The trend of the strength of the quadrupolar component in the combined function dipoles and in the strengths of the chromatic sextupoles is very clear: longer cells allow a reduction of the required $\hat{K}_2$ and $\hat{K}_3$ as does changing the phase advance from $90^\circ$ degrees to $60^\circ$ degrees. Although it is not in the scope of this article to perform a full optimisation of the proposed combined-function periodic cell, these considerations are nevertheless relevant to guide future studies in view of an optimised cell layout. 
\section{Superconducting magnets for a combined-function cell} \label{sec:magnets}
\begin{figure}[htb]
\centering
\includegraphics[trim= 2mm 4mm 2mm 4mm,width=0.4\textwidth,clip=]{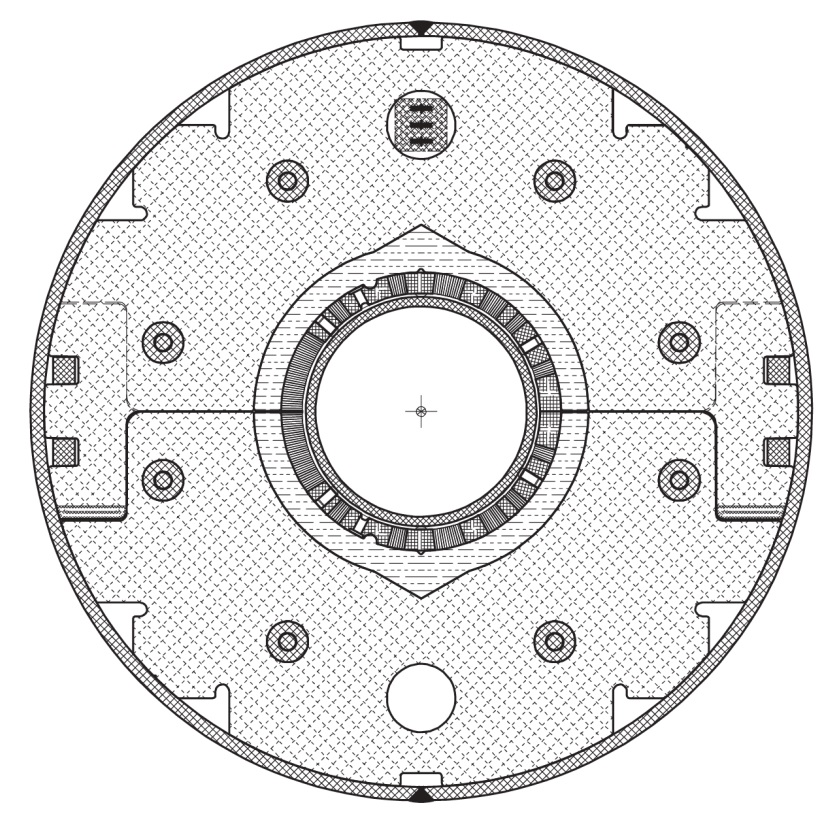}
\includegraphics[trim= 0mm 0mm 0mm 0mm,width=0.53\textwidth,clip=]{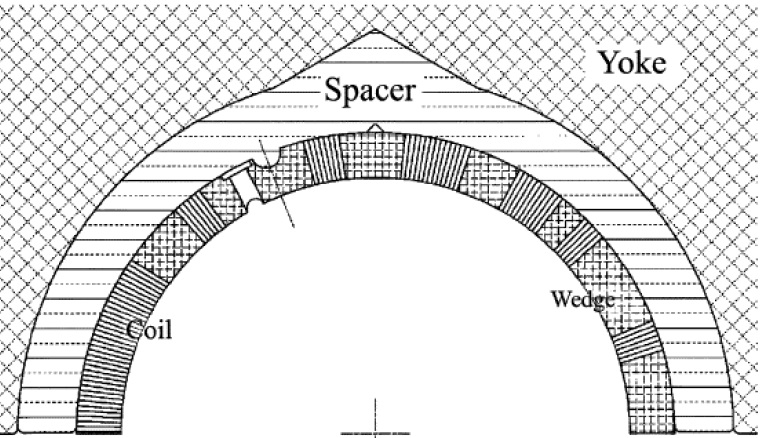}
\caption{Cross section of the single-aperture magnet for the transfer lines of J-PARC (left) and detail of the upper half coil (right) (courtesy T.~Nakamoto).}
\label{fig:J-PARC_magnet}
\end{figure}

Superconducting magnets usually aim at generating a pure multipolar component of the magnetic field, \eg a dipole (for bending), a  quadrupole (for focusing/defocusing), or a sextupole (for chromatic correction). This choice is due to an operational wish, \ie having independent hardware providing each functional requirement. However, from the point of view of design, nothing prevents one from building a magnet that provides several components of field harmonics at the same time. This is routinely done for low-energy machines, relying on resistive magnets, where combined-function magnets can represent a considerable simplification and cost saving. There are also a few cases of combined-function superconducting magnets. For example, in the J-PARC~\cite{Yamazaki:2003wr,Nakamoto2005,Ogitsu2005} transfer line, superconducting magnets providing both a dipolar ($2.6$~T) and a quadrupolar component ($19$~T/m) have been successfully built. Their cross section and details of the coil are shown in Fig.~\ref{fig:J-PARC_magnet}.

The simplest way to compare the two components is to compute them at the aperture radius. For the dipole, the value is the same as that in the centre of the magnet, while for the quadrupole the value is the gradient times the aperture radius. For example, in the J-PARC magnets~\cite{Nakamoto2005,Ogitsu2005} having a $173$~mm aperture diameter, the quadrupolar component gives a peak field at the bore aperture of $1.64$~T, \ie smaller than the dipolar component of $2.6$~T, but of the same order of magnitude. For this reason, the coil is strongly left-right asymmetric, as shown in Fig.~\ref{fig:J-PARC_magnet}. An additional constraint is given by the fact that a superconducting magnet is limited by the combination of the current density and peak field in the coil; adding a quadrupolar component to a dipole will increase the peak field $B_{\rm p}$ in the coil by a quantity that is proportional to the gradient $G$ times the aperture radius $r$:
\begin{equation}
   \Delta B_{\rm p} = \lambda G r  \, , 
\end{equation}
where the parameter $\lambda$ depends on the ratio between the coil width $w$ and the aperture radius according to (see~\cite{PhysRevSTAB.9.102401})
\begin{equation}
   \lambda = a_{-1} \frac{r}{w} + 1 + a_{1} \frac{w}{r} \, , 
\end{equation}
and $a_1$ = 0.11, $a_{-1}$ = 0.042. In the case of the J-PARC combined-function dipole, the coil width is $15$~mm, aperture radius is $86$~mm and therefore $\lambda = 1.25$, \ie there is a $25$\% overshoot of the quadrupolar field in the coil with respect to the product $G \, r$. Therefore, the peak field in the coil is not simply the peak field of the dipole increased by $1.64$~T, but it is the peak field of the dipole increased by $2.05$~T. In our example of a combined function lattice of a hadron collider, the quadrupolar component is obviously a second-order effect with respect to the main field. Among the cases shown in Section~\ref{sec:layout}, if we consider the $90^\circ$ degrees phase advance cell, having the same beta function as the FCC-hh baseline lattice, this corresponds to adding a quadrupolar component of $28.8$~T/m in a $16$~T magnet with a $50$~mm diameter aperture. This provides an additional field at the edge of the aperture of $0.72$~T, \ie twenty times less than the dipolar field. The overshoot can be smaller than in the J-PARC magnet since the larger ratio between coil width and aperture radius allows to have a $\lambda$ as small as 1.15. Therefore, only $0.80$~T are removed from the dipole in terms of superconductor efficiency, \ie the dipole field has to be reduced from $16$~T to $15.2$~T. However, since the quadrupole magnets are removed from the layout, this allows recovering about $7$~m (magnetic length of $6.4$~m plus the magnet heads) every 100 m of half cell, \ie $7$\% of filling factor. Assuming an unchanged beam energy, we can therefore reduce the field of the dipoles to $14.8$~T, have a peak field in the combined-function dipole still lower than the peak field in the $16$~T dipole, and remove the need of building cell lattice quadrupoles. 

\begin{figure}[htb]
\centering
\includegraphics[trim= 2mm 2mm 0mm 1mm,width=0.45\textwidth,clip=]{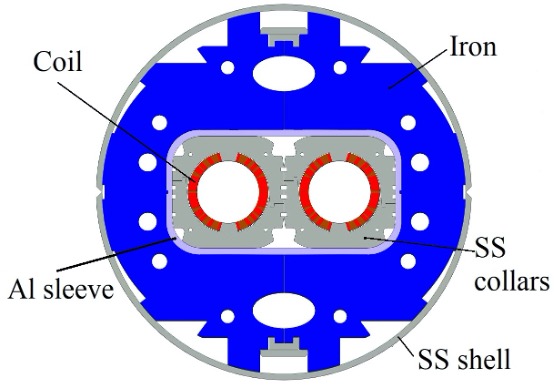}
\includegraphics[trim= 0mm 2mm 2mm 20mm,width=0.53\textwidth,clip=]{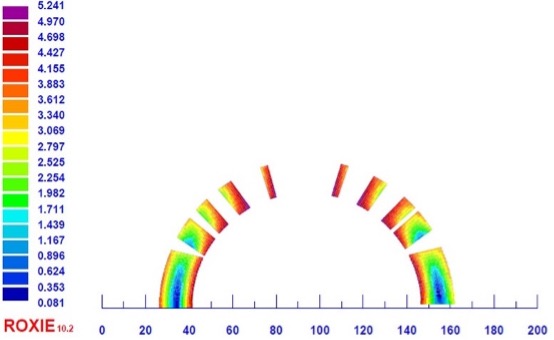}
\caption{Cross section of the double-aperture recombination dipole D2 in HL-LHC (left) and detail of the upper half coil of the left aperture of D2 (right).}
\label{fig:magnet}
\end{figure}

Coil layouts giving a dipolar field as the main component and having a second-order quadrupolar component were also proposed for separation dipoles for future accelerators~\cite{Zlobin2007}. In these double-aperture magnets, the distance between the beams can be similar to the aperture radius, and therefore a considerable magnetic cross-talk between the two bores occurs. In order to remove this component, it was proposed to have the coils slightly left-right asymmetric, to create, in a stand-alone configuration, the same quadrupolar component due to magnetic coupling, but with the opposite sign. Since the quadrupolar component is in this case of the order of a few percent of the main field, the coil is just slightly (\ie order of $1$~mm) left-right asymmetric. This solution has been adopted in the case of the D2 HL-LHC separation  dipole~\cite{BejarAlonso:2749422} (see Fig.~\ref{fig:magnet} showing the double-aperture magnet cross section (left), and the asymmetric coil (right)), providing a dipolar field of $4.5$~T and having a gradient times the aperture radius of 0.135~T~\cite{Farinon2016}. A short-model magnet has been built in 2020, and magnetic measurements validated the design of the asymmetric coil. 
\section{Final considerations on the proposed combined-function periodic cell} \label{sec:opt}
The discussions carried out in the previous sections show that replacing the present FCC-hh cell with a combined-function periodic cell would need a diffused quadrupole gradient of $51.2$~T/m, requiring adding $1.2$~T to the peak field of the dipole. This represents $7.5$\% of the $16$~T dipolar field, and is close to the increase of the filling factor, corresponding to the removal of the $7$~m-long quadrupoles each $100$~m of the FODO half cell. In this configuration, one would end up with a magnet requiring the same peak field, but with the main advantage of having removed all cell quadrupoles.

There is an additional positive side effect that should be taken into account: the maximum beta function in the combined function cell is $20$\% lower than the corresponding FODO cell. Therefore, a fair comparison should involve lattices with the same peak beta function, which increases the combined-function cell length from $188$~m to $250$~m, as shown in Table~\ref{table:summary}. In this case the requirement on the gradient drops by $25$\%, allowing one to recover another $2$\% of integrated dipole field, that could be used to reduce the main dipolar field, always at constant beam energy. 

A third advantage can be gained by spreading the lattice correctors (chromatic, orbit, tune, and Landau octupoles) in the dipoles. In this case, one could remove the need of a cryostat (and a 1-m-long interconnection), plus the dedicated space needed for these correctors. This possibility will be explored in a future paper and could bring another few percent increase in the filling factor. The estimates derived for the chromatic correction show that the combined-function lattice with longer cell, and with the same beta peak as the nominal FCC-hh FODO cell, would require about $30$\% lower strength.   
\section{Conclusions} \label{sec:conclusions}
In this paper, the analysis of a periodic combined-function cell for a  high-energy circular storage ring or collider has been presented. The performance of this periodic lattice structure has been considered in detail, performing an in-depth comparison with the standard FODO cell. For the same cell length, deflection angle, and phase advance, the combined-function structure features smaller values of the beta and dispersion function, which is an advantage in terms of the needs of the beam aperture. Particular attention has been paid to the performance of the periodic cell in terms of chromatic properties and their compensation. The analyses carried out show that a mild increase of the strength of the chromatic sextupoles has to be expected for the combined function cell with respect to the reference structure, \ie the FODO cell. Nonetheless, thanks to the smaller values of the optical functions, the resonant driving terms of the combined function cell are almost the same at those of the FODO cell in the the interval $\muc/2\pi < 0.3$ that covers the needs of most applications. It is also worth mentioning that the change of momentum compaction factor for the combined function cell with respect to the FODO one is very small and no major impact on the longitudinal beam dynamics has to be expected.

In view of finding a periodic lattice structure with favourable chromatic properties, a variant of the combined-function cell has been studied in detail. Such a structure is based on an asymmetric arrangement of the magnets, so that the lengths of the focusing and defocusing blocks of magnets are not equal. Although this configuration has some interesting features, it does not bring any reduction in terms of strength of the chromatic sextupoles and has been disregarded. 

The main advantage of the combined-function cell with respect to the FODO one is the possibility of eliminating the quadrupoles for the arcs of a circular storage ring or collider, which is a considerable simplification. For the considered cell length, the larger filling factor due to the removal of the quadrupoles perfectly compensates the peak field increase required in the dipole magnet. Therefore, the combined-function cell layout is neutral is terms of energy versus dipole field. 

Some variants of the proposed combined-function cell have been considered, in which the number of dipoles and the phase advance have been varied. The expected  trends in terms of optical parameters are confirmed, \ie that the  strength of the quadrupolar component in the combined function elements as well as of the chromatic sextupoles scale with the inverse of the cell length. This information will be crucial in the process of optimisation of the cell layout. 

Since the combined-function cell option provides smaller peak beta-function values, one can increase the cell length so to carry out a fair comparison, in terms of optical parameters, with the FODO lattice. In this way, the quadrupole gradient can be reduced, which provides another $2$\% increase of the integrated dipole field that can be used to reduce the dipole field, for the same proton energy.

Another interesting option, not explored in this paper, is to distribute the lattice correctors in the dipoles. This would allow to further increase the filling factor of few more percent. Among the possible use of this additional margin, \ie to reduce the dipole field or to reduce the overall ring length at constant energy, or to increase further the beam energy at constant ring circumference, the first option appears to be the more attractive one since the marginal cost of the dipole field in the $15-16$~T region is extremely high and nonlinear. In the scenario of a complete removal of the short straight section (both quadrupole and the various magnetic correctors), one also would end up with only one cold mass type in the arcs, namely the combined-function dipole plus a corrector, with an important simplification of the ring design and the industrial process required to build, which, in the end, may lead to a non-negligible cost reduction.
\section{acknowledgement}
We would like to express our warm gratitude to A.~Devred for the suggestion of exploring this option for FCC-hh, T.~Nakamoto for valuable help in describing the J-PARC magnet, and P.~Fabbricatore and S.~Farinon for the magnetic design of the D2 magnet. We would also like to express our warm thanks to R. Tom\'as for interesting discussions and the careful feedback on the original version of the manuscript.
\clearpage
\appendix
\section{Collection of beam dynamics properties of a FODO-cell structure} \label{sec:app1}
In this appendix, some useful properties of the optical parameters of a FODO cell have been collected. The main hypothesis is that the quadrupoles are assumed to be represented as thin lenses. For the sake of simplicity, the properties are assumed to be the same for the two transverse planes, \ie the phase advance is assumed to be the same, but generalisations to the case of different phase advances are straightforward. The cell bending radius is given by $\rc \sin \phc/2=\lc/2$ or $\rc \approx \lc/\phc$. Obviously, one has that $Q=\muc/(2\pi)$.

\begin{figure}[htb]
\centering
\includegraphics[width=0.49\textwidth]{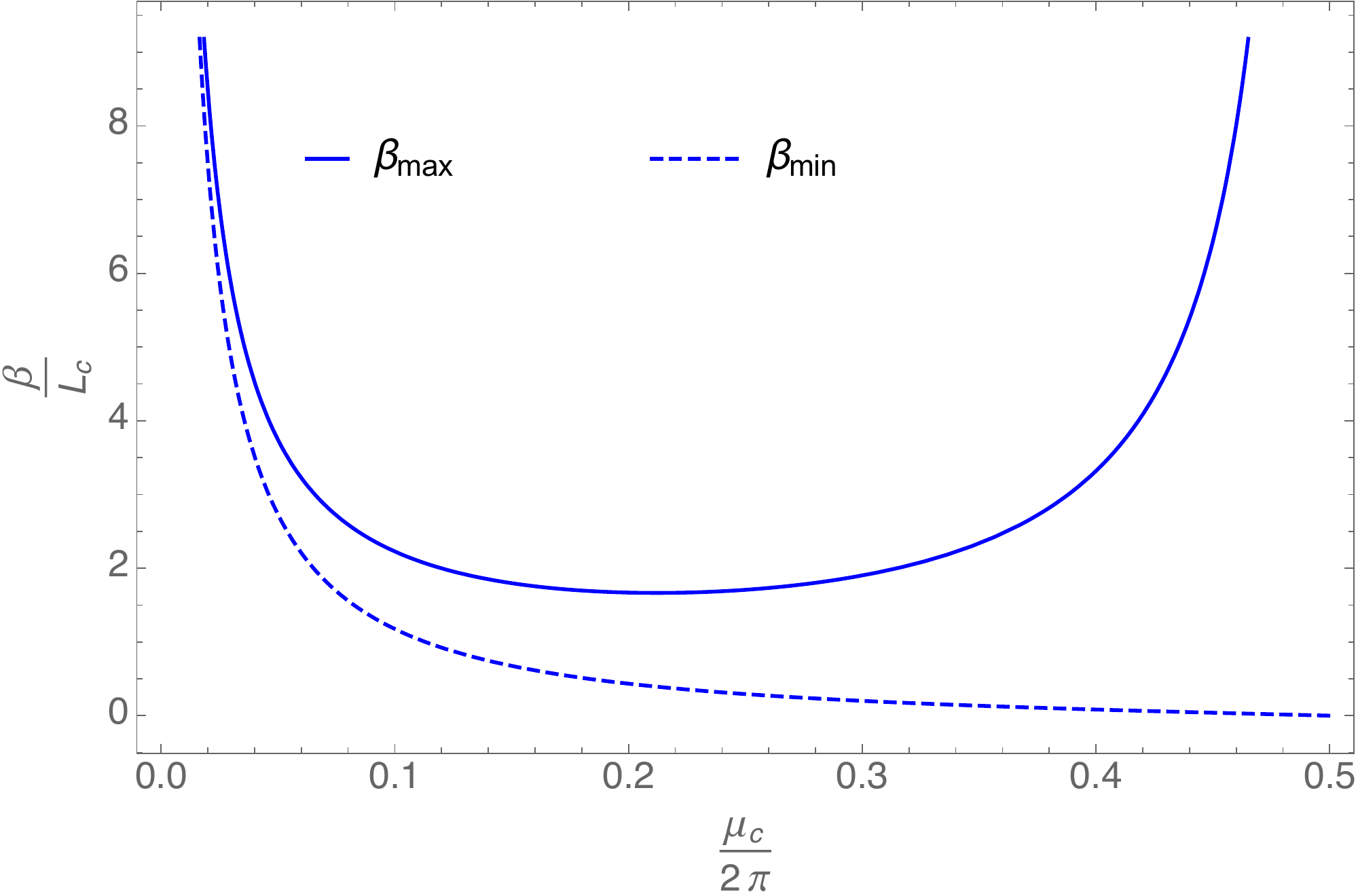}
\includegraphics[width=0.49\textwidth]{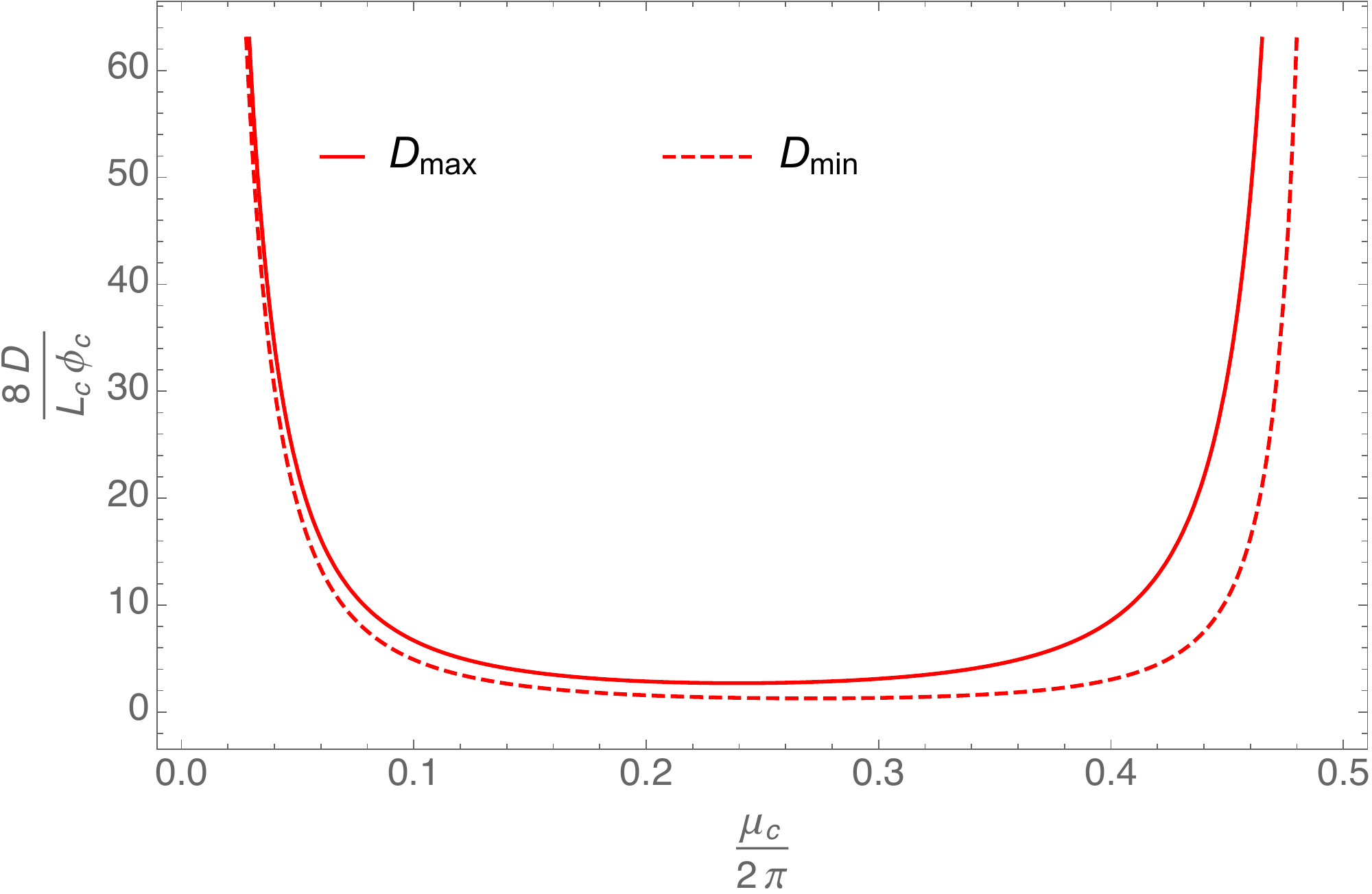}
\caption{Behaviour of the $\beta$- and dispersion-function in dependence of $\muc$.}
\label{fig:FODO}
\end{figure}

The main relations are found for the minimum and maximum values for the $\beta$- and dispersion-function, namely
\begin{eqnarray}
\bma & =\displaystyle{\frac{1+\sin \much}{\sin \muc}}\, \lc \qquad \qquad \dma & = \displaystyle{\frac{2+ \sin \muc}{\sin^2 \muc}} \displaystyle{\frac{\lc \phc }{8}} \\
\bmi & =\displaystyle{\frac{1-\sin \much}{\sin \muc}}\, \lc \qquad \qquad \dmi & = \displaystyle{\frac{2- \sin \muc}{\sin^2 \muc}} \displaystyle{\frac{\lc \phc }{8}} \, ,
\end{eqnarray}
from which the scaling properties of the geometrical characteristics of the FODO cell, \ie total length and bending angle, are explicitly shown, and the plots are shown in Fig.~\ref{fig:FODO}. Note the expressions for $\dma$ and $\dmi$ are based on the assumption that all the cell is filled with a dipolar field, which is consistent with the thin-lens approximation of the quadrupoles.  

While $\bmi$ does not feature a maximum or minimum, the other functions do and it is possible to find the special values of the phase advance for which the minimum is reached and this is given by
\begin{equation}
\begin{split}
    \frac{\dd \bma}{\dd \muc} & =0 \quad \Rightarrow \quad \frac{\muc}{2\pi}= \frac{1}{\pi} \arctan \sqrt{\frac{\sqrt{5}-1}{2}} \approx 0.2121 \\
    \frac{\dd \dma}{\dd \muc} & =0 \quad \Rightarrow \quad \frac{\muc}{2\pi} \approx 0.2392 \\
    \frac{\dd \dmi}{\dd \muc} & =0 \quad \Rightarrow \quad \frac{\muc}{2\pi} \approx 0.2709 \, ,
\end{split}
\end{equation}
so that the minimum of $\dma$ and $\dmi$ is achieved close to $\muc/(2\pi)=1/4$ and the minimum of $\bma$ is only a little further away. In this respect, the value $\muc/(2\pi)=1/4$ provides a close-to-optimal value of the optical and dispersion parameters, in addition to other advantages that are not relevant in this discussion. 

The expressions for the ratio of the maximum and minimum values can be easily derived, giving
\begin{eqnarray}
\frac{\bma}{\bmi} & =\displaystyle{\frac{1+\sin \much}{1-\sin \much}} \qquad \qquad \frac{\dma}{\dmi} & = \displaystyle{\frac{2+ \sin \muc}{2- \sin \muc}} = \displaystyle{\frac{3\bma+\phantom{3} \bmi}{\phantom{3} \bma+3\bmi}} \, .
\label{eq:beta_disp}
\end{eqnarray}

The chromaticity can be obtained starting from the fundamental relationship linking the phase advance to the quadrupole integrated gradient, namely
\begin{equation}
\frac{1}{4} \mathcal{K} \, \lc = \sin \much \qquad \text{where} \qquad \mathcal{K}=\frac{\lqa}{B\rc} \frac{\partial B_y}{\partial x}=\frac{\lqa \, G}{B\rc} \, ,
\label{eq:grad_mu}
\end{equation}
and $\lqa$ represents the length of the quadrupoles and $G$ the magnetic gradient and $\mathcal{K}$ is the integrated normalised gradient. Considering that in the presence of a momentum deviation $\Delta p/p$, the following transformations are applicable $K \to K(1-\Delta p/p)$ and $\muc \to \muc + \mucp \, \Delta p/p$, so that from Eq.~\eqref{eq:chrom} one obtains
\begin{equation}
\mucp = -2 \tan \much \qquad \text{or} \qquad Q'=\frac{\mucp}{2\pi}=-\frac{1}{\pi} \tan \much \, .
\label{eq:chrom}
\end{equation}
From the very definition, it is possible to find an expression for the momentum compaction $\alp$, \ie
\begin{equation}
\alp=\frac{1}{\lc} \int_0^{\lc} \frac{D(s)}{\rho(s)} d s = \frac{\phc}{\lcs} \int_0^{\lc} D(s) d s = \phcs \, \frac{8-\sin^2}{32 \sin^2 \much} \, . 
\label{eq:mom_comp}
\end{equation}

The correction of the linear chromaticity in a FODO cell can be performed by means of a pair of sextupoles placed at the location of the two quadrupoles (represented in the thin-lens approximation). Let us indicate with $\ksf, \ksd$ the integrated strength of the sextupoles located next to the focusing and defocusing quadrupoles, respectively. Then, it is possible to write~\cite{bryant_johnsen_1993}
\begin{equation}
    \begin{split}
        \Delta Q'_x & = -\frac{1}{4\pi} \left ( \Delta \ksf \bma \dma +\Delta \ksd \bmi \dmi \right ) \\
        \Delta Q'_y & = \phantom{-} \frac{1}{4\pi} \left ( \Delta \ksf \bmi \dma +\Delta \ksd \bma \dmi \right ) \, ,
    \end{split}
    \label{eq:chromcorr0}
\end{equation}
from which the solution for the strength of the chromatic sextupoles is given by
\begin{equation}
    \begin{split}
        \ksf & = \phantom{-}\frac{4\pi}{\dma} \frac{ \bma \Delta Q'_x + \bmi \Delta Q'_y}{\bmi^2-\bma^2} \\
        \ksd & = -\frac{4\pi}{\dmi} \frac{ \bmi \Delta Q'_x + \bma \Delta Q'_y}{\bmi^2-\bma^2} \, .
    \end{split}
    \label{eq:chrom_corr1}
\end{equation}

Note the quantities in the denominator are pushing the strength of the sextupole up (in absolute value) if the difference between the maximum and minimum values of the beta function is reduced. A similar effect is observed on the value of $\ksf$, whenever $\dma$ is reduced.

\begin{figure}[htb]
\centering
\includegraphics[width=0.59\textwidth]{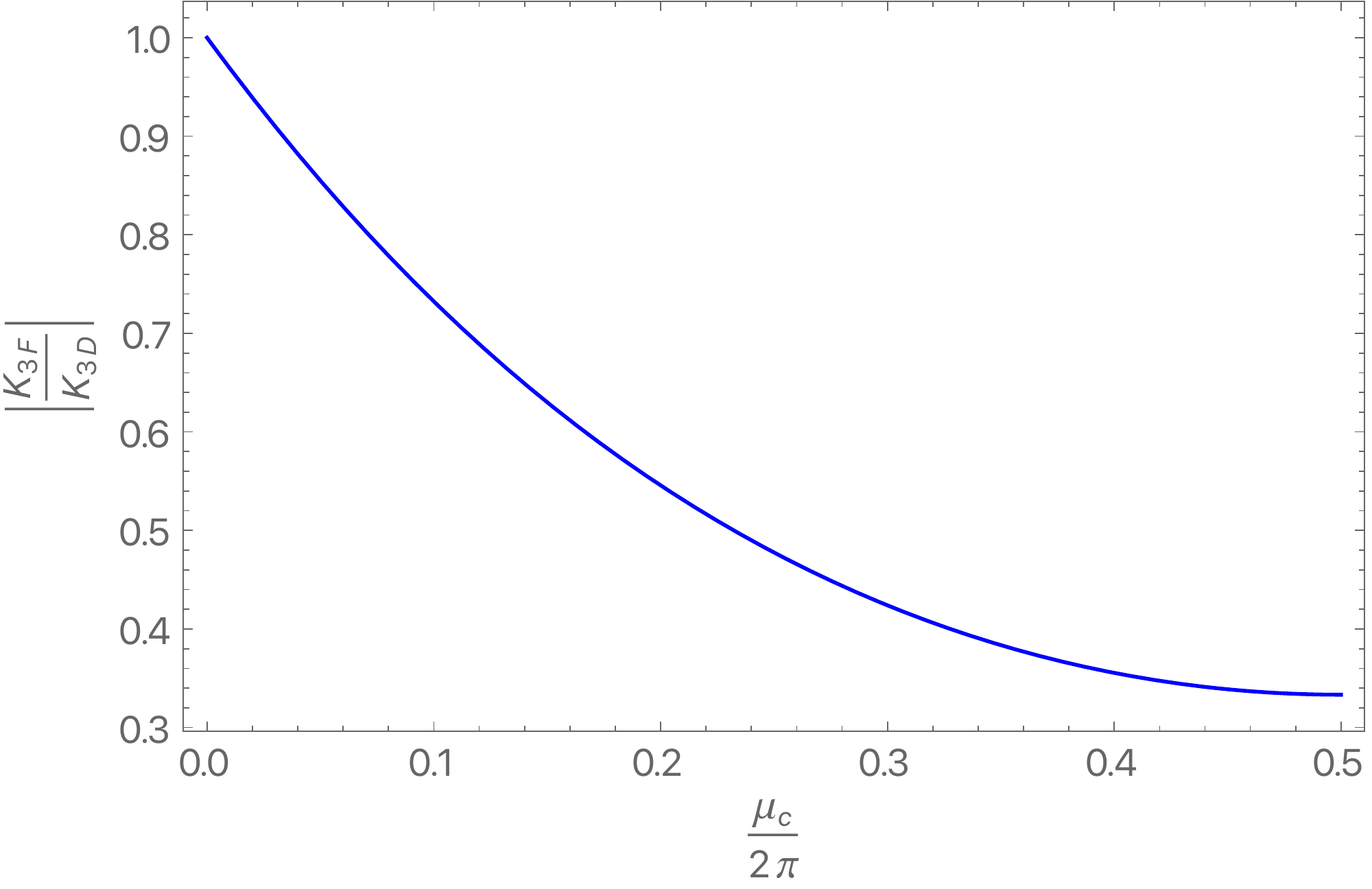}
\caption{Behaviour of the ratio of the strength of the sextupoles used to correct the chromaticity of a FODO cell in dependence of $\muc$.}
\label{fig:sexratio}
\end{figure}

By means of the relationships established between the optical parameters and the properties of the FODO cell, \ie phase advance $\muc$ and bending angle $\phc$, it is possible to recast Eq.~\eqref{eq:chrom_corr1} in the following form
\begin{equation}
    \begin{split}
        \ksf & = -\frac{1}{\lc^2 \phc} \frac{\sin^2 \much}{\cos \much} \left ( \frac{\sin \much+1}{\sin \much +2} \Delta Q'_x - \frac{\sin \much -1}{\sin \much +2 } \Delta Q'_y\right ) \\
        \ksd & = \phantom{-}\frac{1}{\lc^2 \phc} \frac{\sin^2 \much}{\cos \much} \left ( \frac{\sin \much-1}{\sin \much -2} \Delta Q'_x - \frac{\sin \much +1}{\sin \much -2 } \Delta Q'_y\right )  \, ,
    \end{split}
    \label{eq:chrom_corr2}
\end{equation}
which can be further simplified, whenever $\Delta Q'_x= \Delta Q'_y=\Delta Q'$, to give
\begin{equation}
    \begin{split}
        \ksf & = -\frac{2}{\lc^2 \phc} \frac{\sin^2 \much}{\cos \much \left ( \sin \much +2 \right )} \Delta Q'\\
        \ksd & = -\frac{2}{\lc^2 \phc} \frac{\sin^2 \much}{\cos \much \left ( \sin \much -2 \right )} \Delta Q'
    \end{split}
    \label{eq:chrom_corr3}
\end{equation}
from which one deduces that 
\begin{equation}
    \frac{\ksf}{\ksd}= \frac{\sin \much -2}{\sin \much +2} 
\end{equation}
and this ratio is plotted in Fig.~\ref{fig:sexratio}. Note for the special value $\muc/(2\pi)=1/4$ the ratio is approximately equal to $1/2$, thus revealing a non-negligible asymmetry between the strength of the two sextupoles.  
\section{Optical properties of a combined-function cell with impact on the longitudinal beam dynamics} \label{sec:app2}
\begin{figure}[htb]
\centering
\includegraphics[trim=0mm 0mm 0mm 0mm, height=0.345\textwidth,clip=]{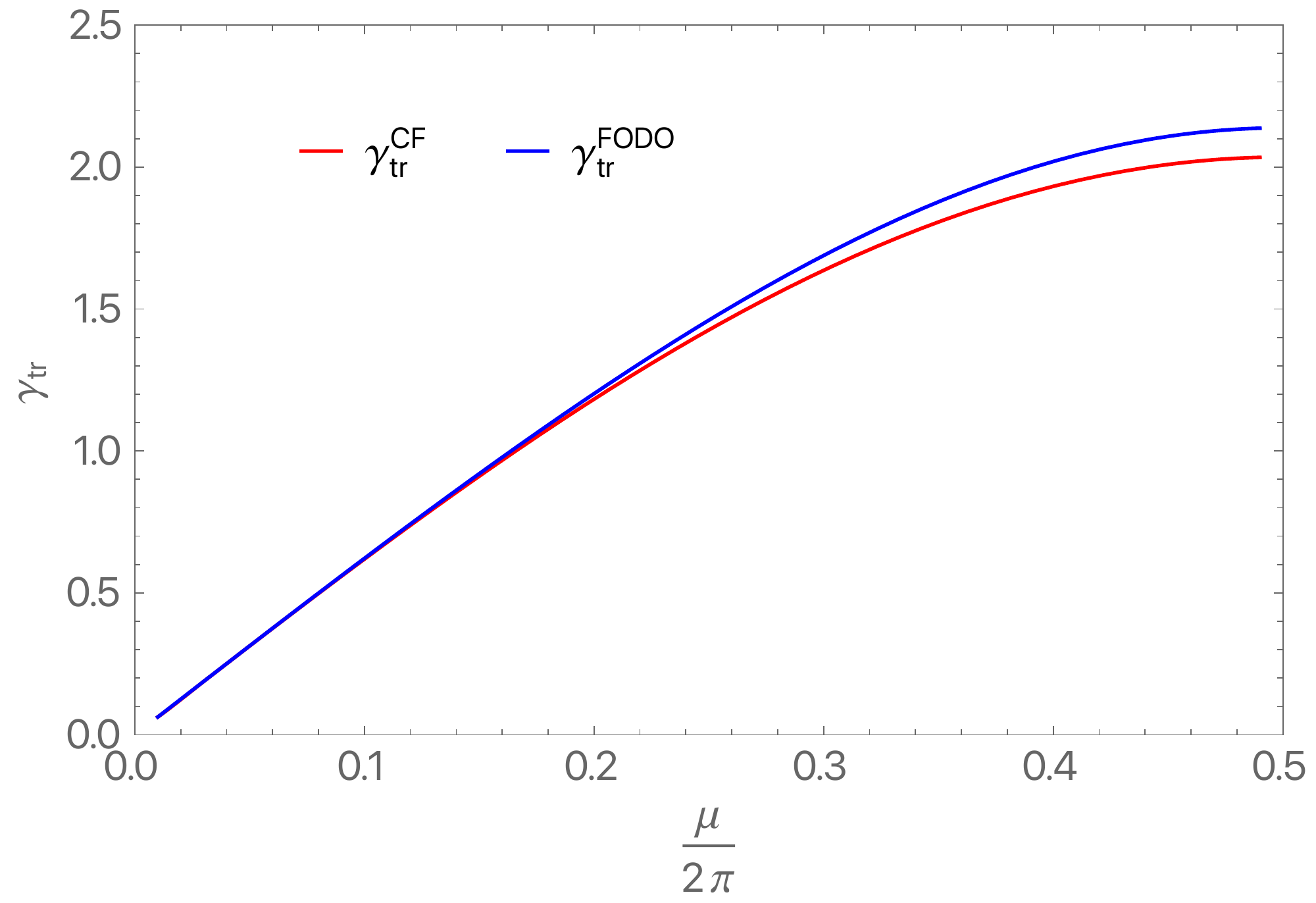}
\includegraphics[trim=0mm 0mm 0mm 0mm, height=0.345\textwidth,clip=]{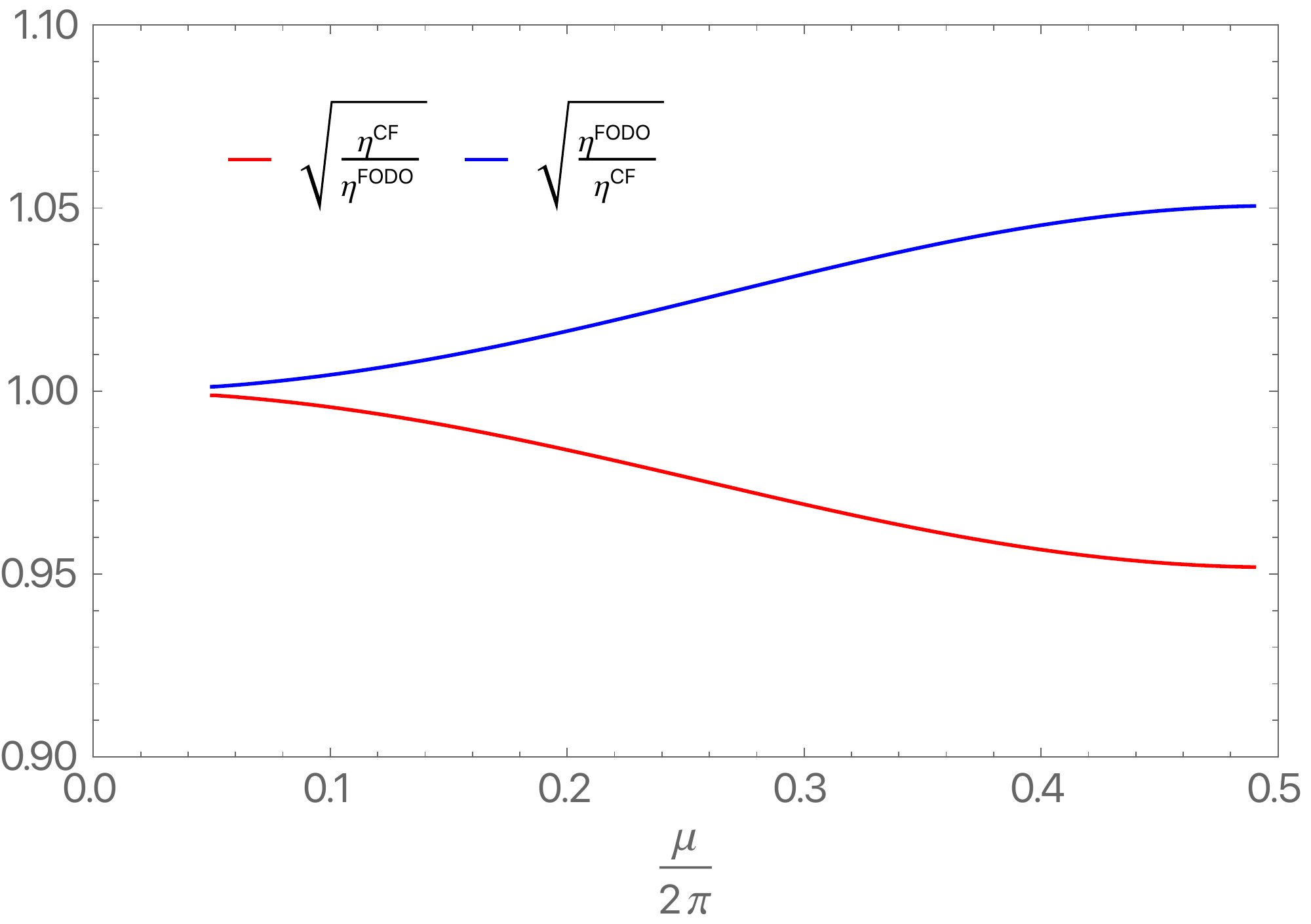}
\caption{Left: $\gtr$ for the FODO and the combined-function cells as a function of the phase advance. Right: ratio of the square root of the $\eta$ factor for the FODO and the combined-function cells as a function of the phase advance.}
\label{fig:rf}
\end{figure}

The optical properties of a periodic cell have an impact on the longitudinal beam dynamics~\cite{Lee:2651939} via the momentum compaction $\alpha_\mathrm{c}$ and the derived quantities, namely
\begin{equation}
    \gtr = \frac{1}{\sqrt{\alpha_\mathrm{c}}} \qquad \eta = \frac{1}{\gamma^2} -\frac{1}{\gtr^2} \, .
\end{equation}
The evolution of $\gtr$ for the two types of periodic cells considered in this paper is shown in Fig.~\ref{fig:rf} (left) as a function of the cell phase advance, and the difference is small over a rather large range of phase advances and increases mildly towards large values of the phase advance. 

To compare the value of the slip factor $\eta$ for the two cells one needs to assume that $\gamma \gg \gtr$, which is certainly correct for high-energy rings, and one obtains
\begin{equation}
    \frac{\eta_\mathrm{CF}}{\eta_\mathrm{FODO}} = \left ( \frac{\gtr^\mathrm{FODO}}{\gtr^\mathrm{CF}} \right )^2 \frac{\left ( \gtr^2-\gamma^2\right )_\mathrm{CF}}{\left ( \gtr^2-\gamma^2\right )_\mathrm{FODO}} \approx \left ( \frac{\gtr^\mathrm{FODO}}{\gtr^\mathrm{CF}} \right )^2 \, .
\end{equation}
Some key quantities for the longitudinal beam dynamics are function of $\eta$, such as 
\begin{equation}
    Q_\mathrm{s} =\sqrt{\frac{h e V_\mathrm{RF} |\eta| | \cos \phi_\mathrm{s}|}{2\pi \beta^2 E}} \qquad \mathcal{A}_\mathrm{b} \propto \sqrt{\frac{e V_\mathrm{RF}}{2\pi \beta^2 E h |\eta|}}
\end{equation}
where $Q_\mathrm{s}$ and $\mathcal{A}_\mathrm{b}$ are the synchrotron tune and the bucket area, respectively. In the previous equations, $e$ is the electron charge, $\phi_\mathrm{s}$ is the synchronous phase, $V_\mathrm{RF}$ the voltage applied by the RF cavities, $h, \beta, E$ are the harmonic number, the relativistic factor, and the beam energy, respectively. Therefore, $Q_\mathrm{s} \propto \sqrt{|\eta|}$, whereas $\mathcal{A}_\mathrm{b} \propto 1/\sqrt{|\eta|}$ and in Fig.~\ref{fig:rf} (right) the ratio of the square root of $\eta$ for the combined function and the FODO cell (or its inverse) is shown as a function of the cell phase advance. 

By considering the typical case $\muc=\pi/2$, one has that the synchrotron tune of a combined function cell will be $2$\% higher than that of a FODO cell, whereas the bucket area will be $2$\% smaller, assuming that $V_\mathrm{RF}$ is the same for both cells. It is of course possible to compensate such a difference by acting on $V_\mathrm{RF}$, which scales linearly with $\eta$ for the case of $\mathcal{A}_\mathrm{b}$ or inversely proportional to $\eta$ for the case of $Q_\mathrm{s}$. In all the cases, the variation of the RF voltage that should be applied is of the order of a few percent, hence negligible. 
\section{Some properties of an asymmetric combined-function cell} \label{sec:app3}
\begin{figure}[htb]
\centering
\includegraphics[width=0.49\textwidth]{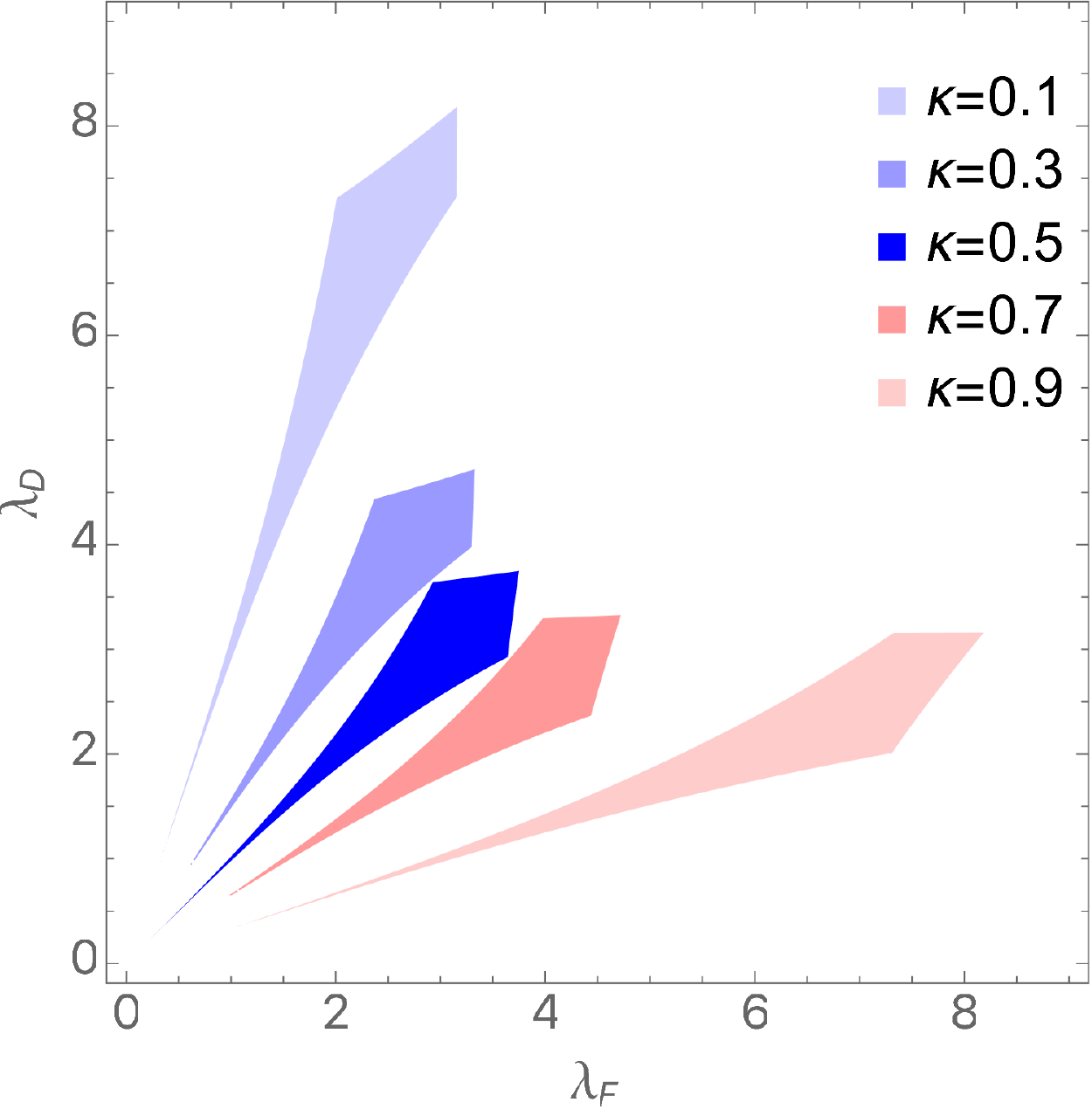}
\caption{Stability region in the $(\laf,\lad)$ space for the asymmetric layout of the combined-function periodic cell as a function of the parameter $\kappa$. The symmetric case, corresponding to $\kappa=0.5$ is also shown for the sake of comparison.}
\label{fig:stabilityasymm}
\end{figure}

In the proposed layout of the combined-function cell, the length of the two blocks of focusing and defocusing magnets is the same. However, it is possible to break this symmetry, considering the case in which $L_\mathrm{F}=(1-\kappa) \, \lc $ and $L_\mathrm{D}=\kappa \, \lc$, with $\kappa \in ]0,1[ $. In this case, the $2\times 2$ cell matrices become
\begin{equation}
\begin{split}
    M_x & = \left(
\begin{array}{cc}
 \frac{\lad^2-\laf^2}{2 \laf \lad}\sinh \ladh \sin \lafh +\cosh \ladh \cos \lafh & -\frac{\lc (\lad^2-\laf^2)}{2 \lad \laf^2} \sinh \ladh
   \left( \cos \lafh -\frac{\lad^2+\laf^2}{\lad^2-\laf^2} \right) +\frac{\lc}{\laf}\cosh \ladh \sin \lafh \\
 \frac{\lad^2-\laf^2}{2 \lc \lad}\sinh \ladh \left( \cos \lafh+\frac{\lad^2+\laf^2}{\lad^2-\laf^2} \right) - \frac{\laf}{\lc} \cosh \ladh \sin \lafh & \frac{\lad^2-\laf^2}{2 \laf \lad}\sinh \ladh \sin \lafh + \cosh \ladh \cos \lafh \\
\end{array}
\right) \\
& \\
M_y & = \left(
\begin{array}{cc}
 -\frac{\lad^2-\laf^2}{2 \laf \lad} \sin \ladh \sinh \lafh + \cos \ladh \cosh \lafh & -\frac{\lc (\lad^2-\laf^2)}{2 \lad \laf^2} \sin \ladh \left( \cosh \lafh - \frac{\lad^2+\laf^2}{\lad^2-\laf^2} \right) + \frac{\lc}{\laf} \cos \ladh \sinh \lafh \\
\frac{\lad}{\lc} \sin \ladh  \left(\frac{\laf^2}{\lad^2} \sinh^2 \laff - \cosh^2 \laff \right ) + \frac{\laf}{\lc} \cos \ladh \sinh \lafh & -\frac{\lad^2-\laf^2}{2 \laf \lad} \sin \ladh \sinh \lafh + \cos \ladh \cosh \lafh \\
\end{array}
\right) \, , 
\end{split}
\label{eq:tmat}
\end{equation}
and the stability conditions read
\begin{equation}
    \begin{split}
      2 \cos \mu_{\mathrm{c},x} & = 2 \cos (1-\kappa)\laf 
      \cosh \kappa \, \lad + \frac{\lad^2 - \laf^2}{\laf \lad} \sin (1-\kappa )\laf \sinh \kappa \, \lad \\  
      2 \cos \mu_{\mathrm{c},y} & = 2 \cos \kappa \, \lad 
      \cosh (1-\kappa) \laf - \frac{\lad^2 - \laf^2}{\laf \lad} \sin \kappa \, \lad \sinh (1-\kappa)\laf  \, ,
    \end{split}
    \label{eq:staba}
\end{equation}
which provide the stability domains shown in Fig.~\ref{fig:stabilityasymm}, where the impact of the additional parameter $\kappa$ introduced in the layout of the combined function cell is clearly visible.

\begin{figure}[htb]
\centering
\includegraphics[height=0.30\textwidth,clip=]{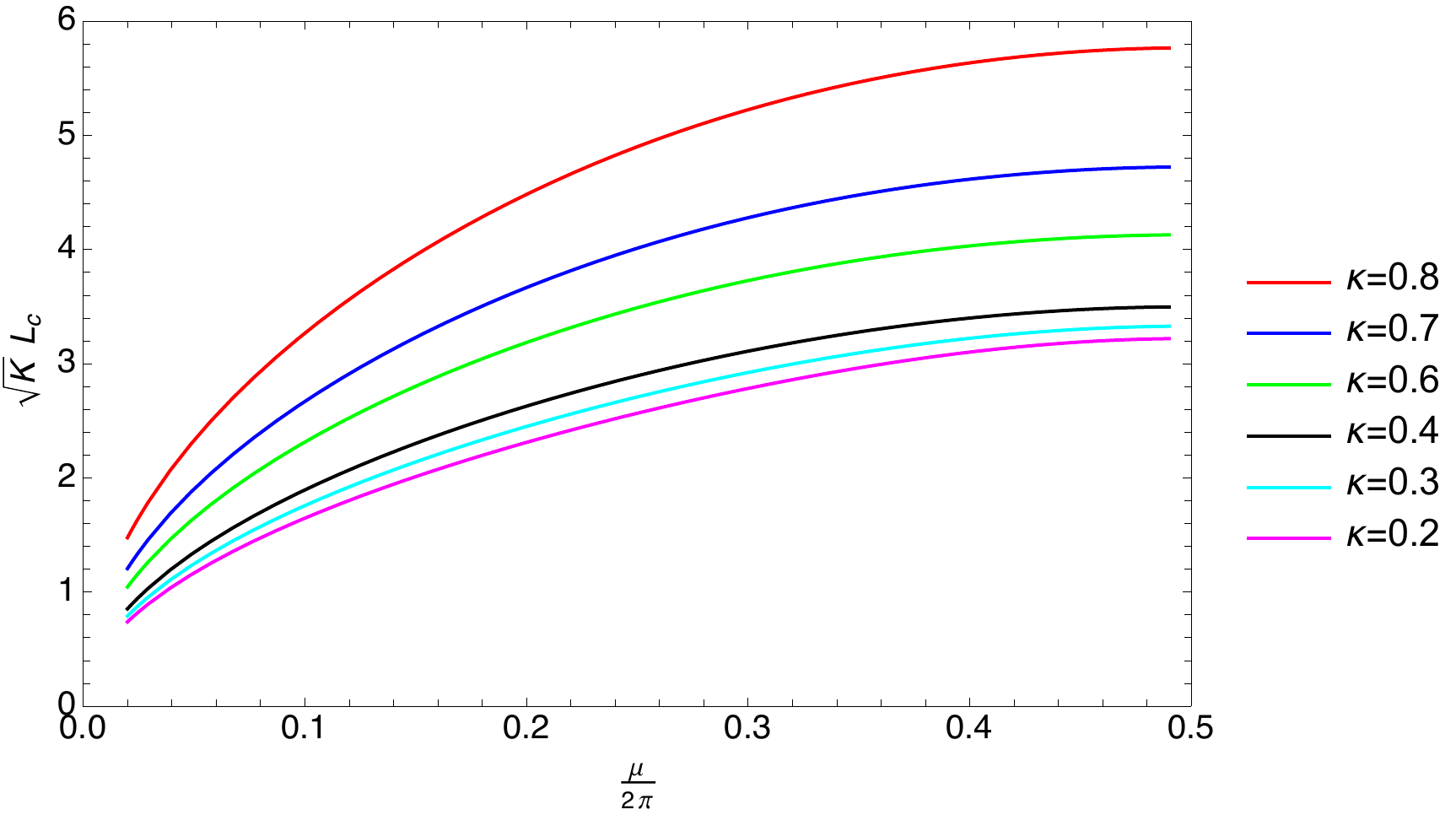}
\includegraphics[trim=0mm 0mm 30mm 0mm, height=0.30\textwidth,clip=]{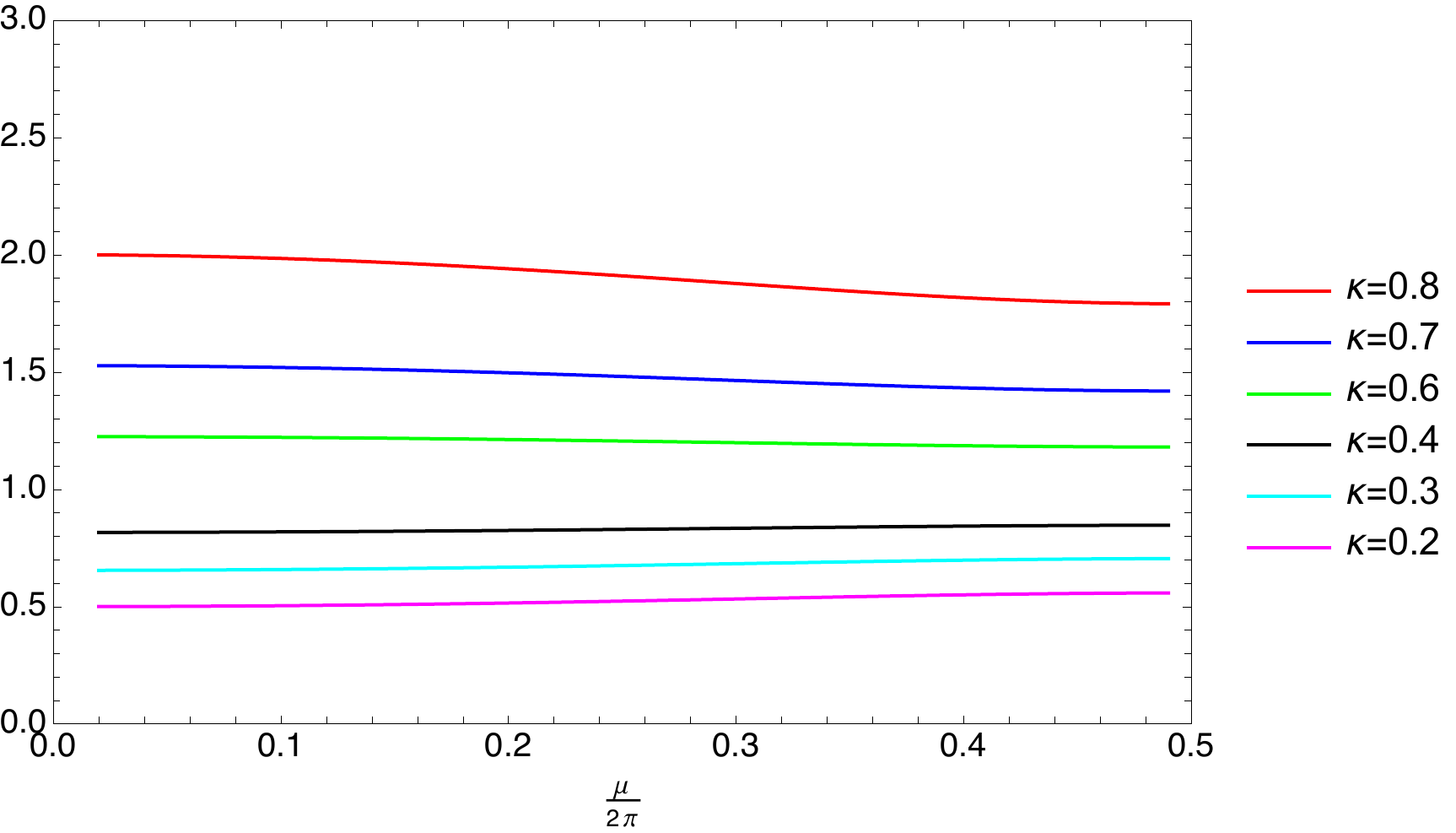}
\caption{Left: solution of Eq.~\eqref{eq:staba}, for the focusing quadrupole strength, as a function of the cell phase advance for different values of $\kappa$. Right: ratio of the strength of the focusing and defocusing quadrupoles.}
\label{fig:gradCFa}
\end{figure}

Equation~\eqref{eq:staba} has to be solved numerically for the integrated gradients $\laf, \lad$ to study the optical parameters of this special configuration of the combined function cell. For the sake of comparison with the simple FODO-cell layout presented in this paper, the condition of having the same tunes in the horizontal and vertical planes is imposed, \ie $\mu_{\mathrm{c},x}=\mu_{\mathrm{c},y}$. This condition represents what is used in a realistic circular accelerator design and, in any case, it would be immediate to obtain solutions for different cell phase advances in the two transverse planes. In Fig.~\ref{fig:gradCFa}, the solution for the focusing quadrupoles strength is shown (left) together with the ratio between the strength of the focusing and defocusing quadrupoles. 

Note that the ratio is nearly constant over the full domain of the phase advance. It is also worth stressing that the solutions of Eq.~\eqref{eq:staba} satisfy the symmetry property $\laf(\kappa)=\lad(1-\kappa)$. These numerically computed solutions are the basis for the study of the optical properties of this asymmetric combined-function cell layout. Analytical expressions for the main optical parameters can be found using standard techniques and some are listed in the following
\begin{equation}
    \begin{split}
        \mu'_{\mathrm{c},x}(\kappa) & = \phantom{-} \frac{\laf \left [ (\kappa +1) \lad^2+(\kappa -1) \laf^2 \right ] \sinh \kappa \, \lad \cos (\kappa -1) \laf-\lad \left [\kappa \lad^2+(\kappa -2) \laf^2 \right ] \cosh \kappa \, \lad \sin (\kappa -1) \laf}{4 \lad \laf} \\
   \mu'_{\mathrm{c},y}(\kappa) & = -\frac{\lad \left [ \kappa \lad^2+(\kappa -2) \laf^2 \right ] \cos \kappa \, \lad \sinh (1-\kappa) \laf+\laf \left [(\kappa +1) \lad^2+(\kappa -1) \laf^2 \right ] \sin \kappa \, \lad \cosh (\kappa -1) \laf}{4 \lad \laf} \\
   D_\mathrm{F}(\kappa) & = -\frac{\laf^2 \sinh \frac{\kappa \, \lad}{2} - \lad^2 \sinh \frac{\kappa \, \lad}{2} \cos \frac{(\kappa-1) \laf}{2} + \laf \lad \cosh \frac{\kappa \, \lad}{2} \sin \frac{(1-\kappa) \laf}{2} + \lad^2 \sinh \frac{\kappa \, \lad}{2}}{\lad \laf^2 \left( \lad \sinh \frac{\kappa \, \lad}{2} \cos \frac{(\kappa-1) \laf}{2} - \laf \cosh \frac{\kappa \, \lad}{2} \sin \frac{(1-\kappa) \laf}{2} \right)} \lc \phc \\
   D_\mathrm{D}(\kappa) & = -\frac{\laf^2 \cosh \frac{\kappa \, \lad}{2} \sin \frac{(\kappa-1) \laf}{2} + \lad \laf \sinh \frac{\kappa \lad}{2} \cos \frac{(\kappa-1) \laf}{2} -\lad^2 \sin \frac{(\kappa-1) \laf}{2} - \laf^2 \sin \frac{(\kappa-1) \laf}{2}}{\lad^2 \laf \left(\lad \sinh \frac{\kappa \, \lad}{2} \cos \frac{(\kappa-1) \laf}{2} - \laf \cosh \frac{\kappa \, \lad}{2} \sin \frac{(1- \kappa) \laf}{2}  \right)} \lc \phc \, .
    \end{split}
\end{equation}

It is interesting to note that, thanks to the symmetry of the solutions of Eq.~\eqref{eq:staba}, it is possible to show that $\mu'_{\mathrm{c},x}(\kappa)=\mu'_{\mathrm{c},y}(1-\kappa)$, whereas no symmetry is present for the case of the dispersion function.

The expressions for the beta-functions are obtained by using the element $(1,2)$ of the transfer matrix of the cell, \eg Eq.~\eqref{eq:tmat} for the evaluation of the beta-function at the entrance of the focusing quadrupole, and dividing it by $\sin \muc$.

\begin{figure}[htb]
\centering
\includegraphics[height=0.30\textwidth,clip=]{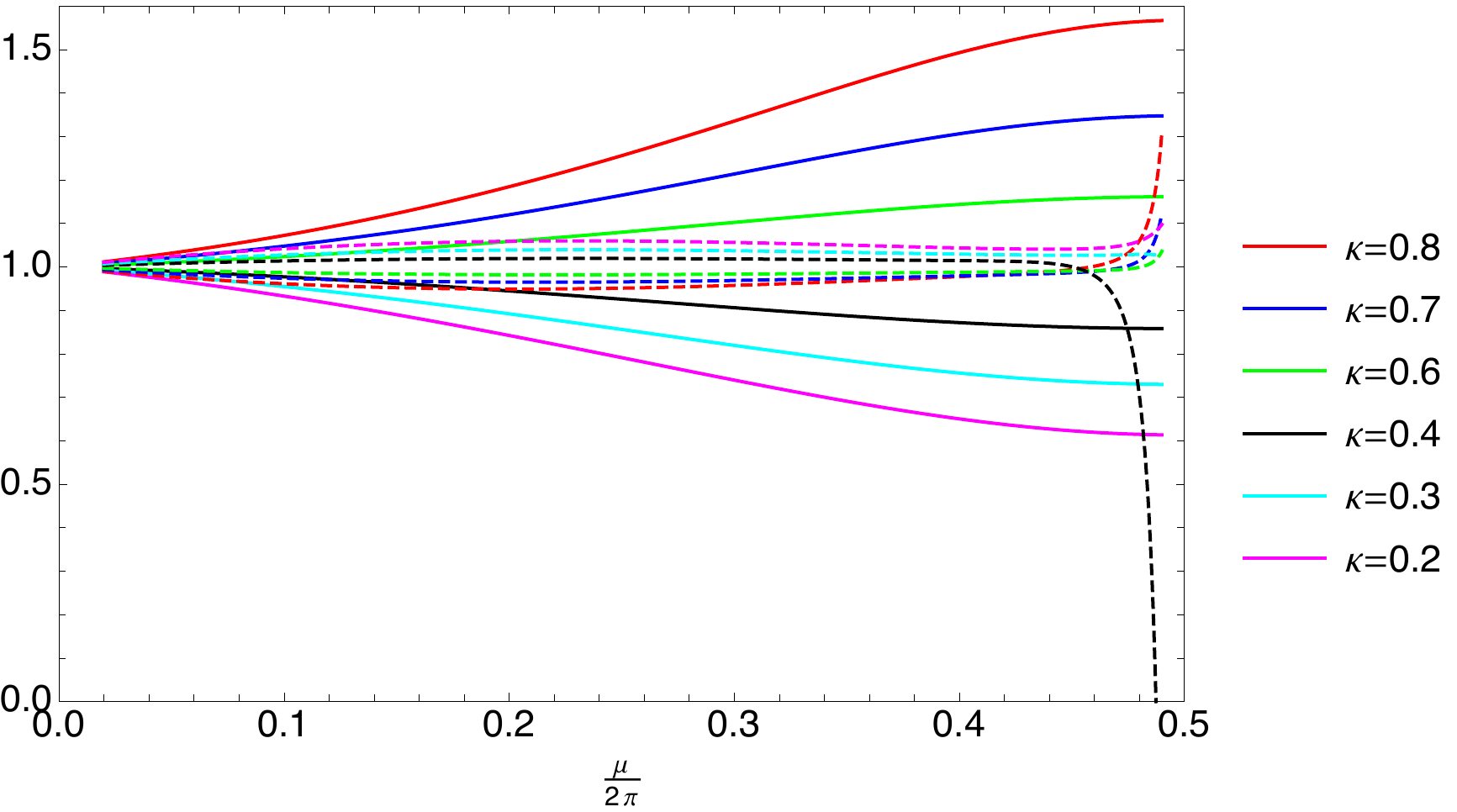}
\includegraphics[trim=0mm 0mm 30mm 0mm, height=0.30\textwidth,clip=]{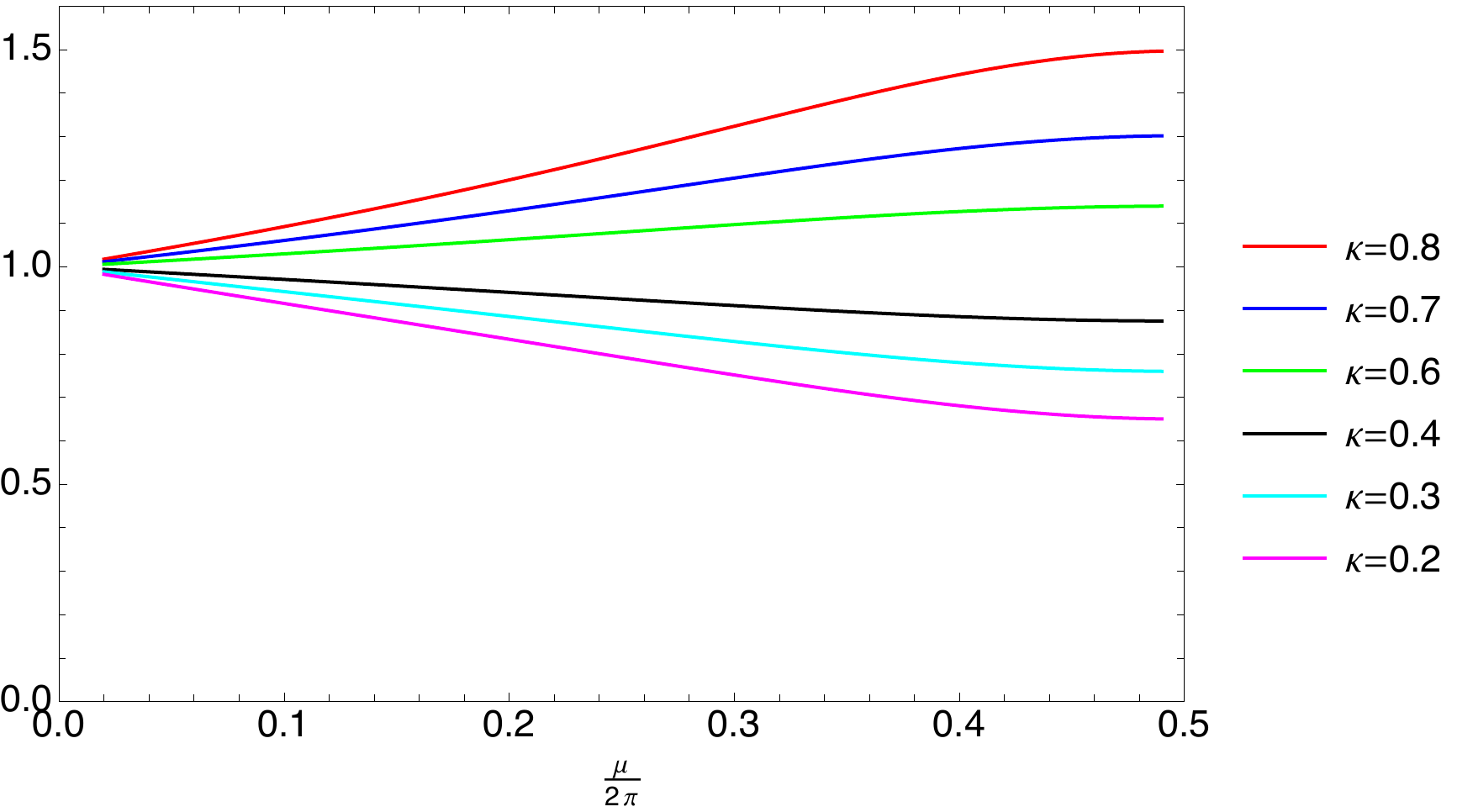}
\includegraphics[height=0.30\textwidth,clip=]{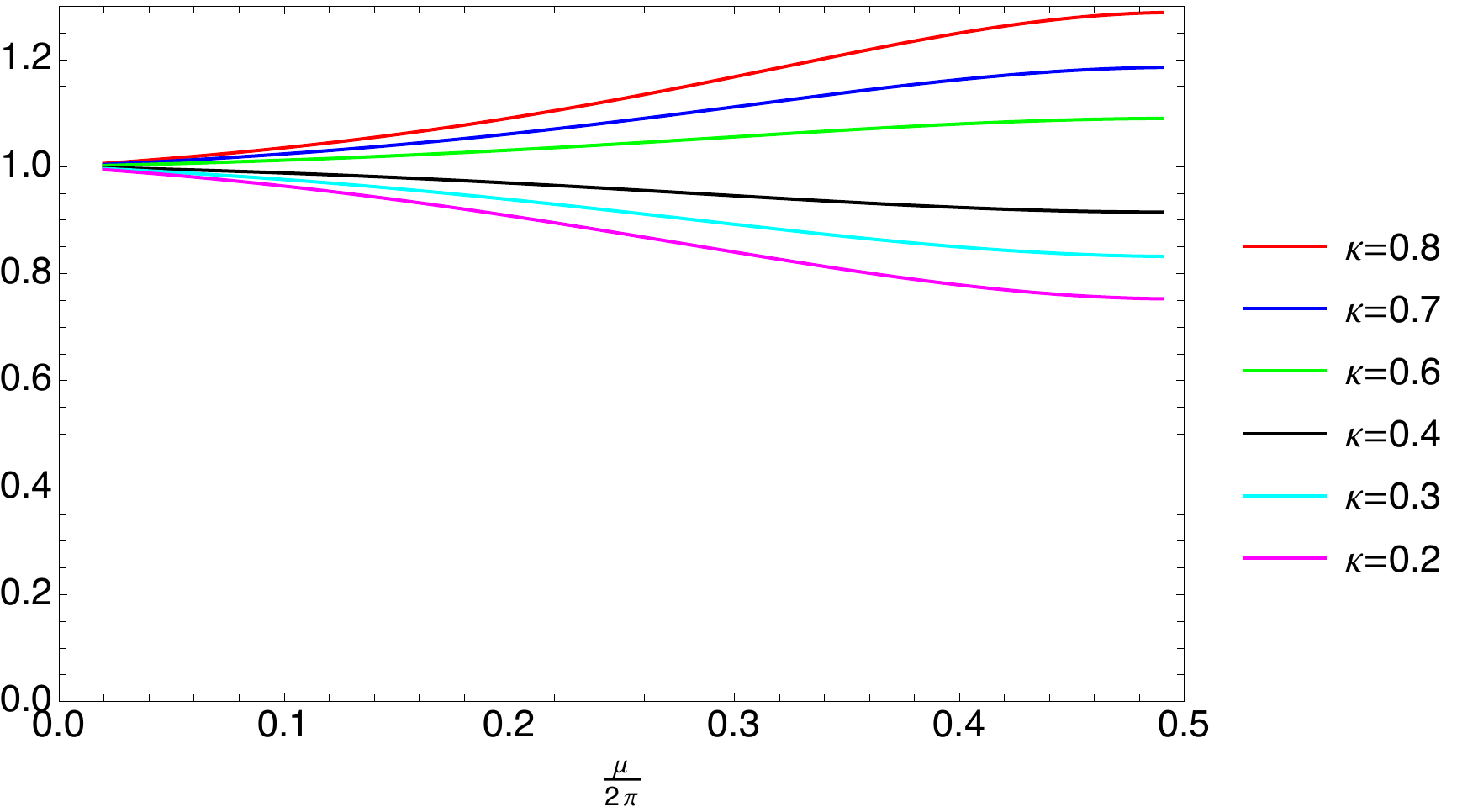}
\includegraphics[trim=0mm 0mm 30mm 0mm, height=0.30\textwidth,clip=]{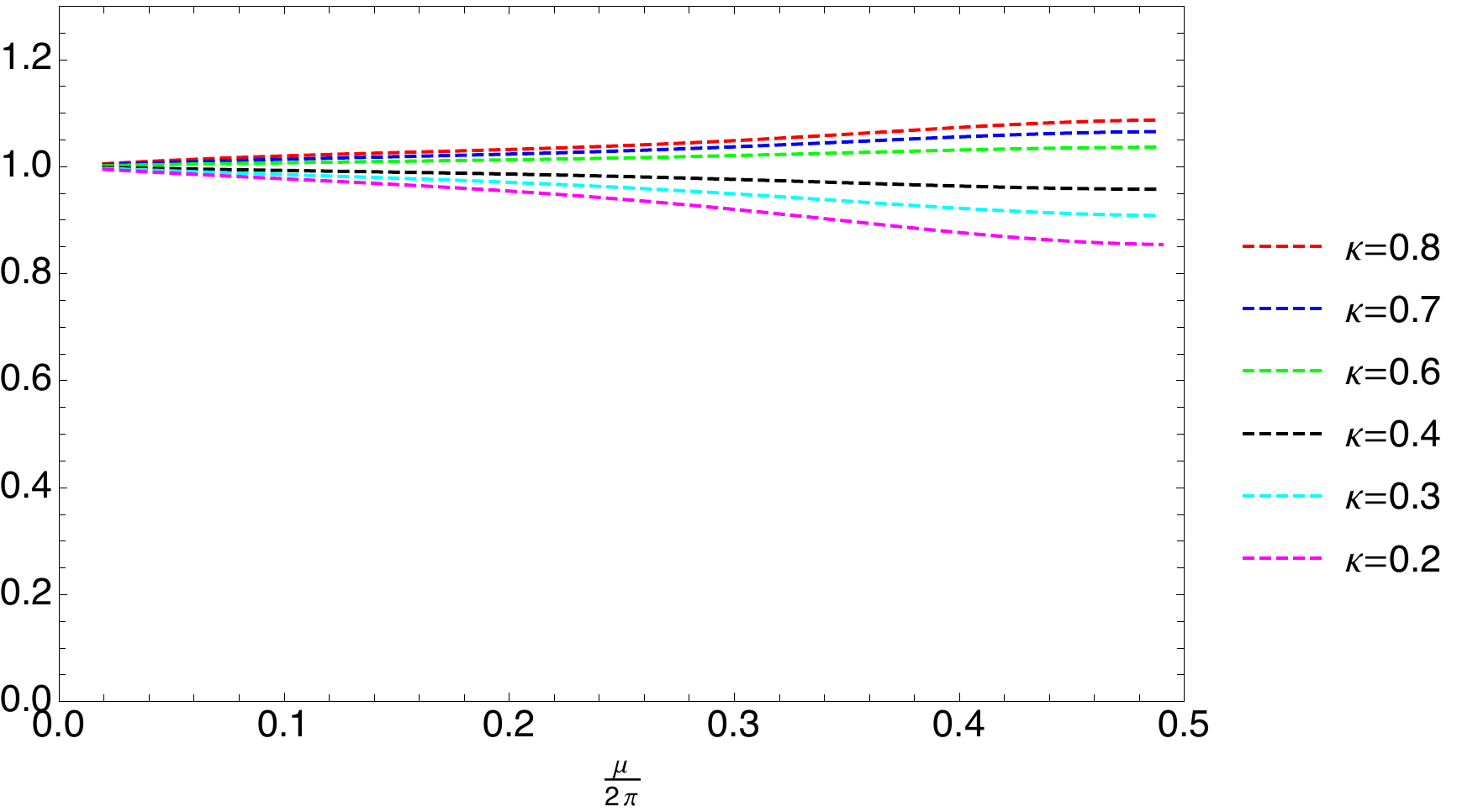}
\caption{Variation of the optical parameters in dependence of $\muc$ for some values of $\kappa$. The ratio between a given optical parameter for an arbitrary value of $\kappa$ and the same parameter for $\kappa=0.5$ is reported. In the upper-left plot $\beta_{\mathrm{F},x}$ (solid lines) and $\beta_{\mathrm{F},y}$ (dashed lines) are shown. In the upper-right plot $Q'_x=\mu'_{\mathrm{c},x}/(2\pi)$ (solid lines) is shown. In the lower plots $D_\mathrm{F}$ (left) and $D_\mathrm{D}$ (right) are shown. Note the symmetry is clearly visible for the case of the beta-function whereas it is lacking for the case of the dispersion function.}
\label{fig:CFpara}
\end{figure}

The variation of the optical parameters as a function of the cell phase advance is visible in Fig.~\ref{fig:CFpara}, where the dependence of the main optical parameters on the cell phase advance is depicted for several values of the parameter $\kappa$. What is shown in the figure is the ratio of the given optical parameter for arbitrary values of $\kappa$ to the same parameter for $\kappa=0.5$, which corresponds to the symmetric combined function layout. The values of $\beta_{\mathrm{F},z}$, $Q'_x=\mu'_{\mathrm{c},x}/(2\pi)$, $D_\mathrm{F}$, and $D_\mathrm{D}$ are shown in the upper-left, upper-right, lower-left, and lower-right plots, respectively.
The existing symmetry, \ie  $\beta_{\mathrm{D},x}(\kappa)=\beta_{\mathrm{F},y}(1-\kappa)$, allows inspecting only the behaviour of the beta function at the focusing quadrupole. Note that such a symmetry is lacking for the dispersion function. One can observe that the parameter $\kappa$ allows a fine-tuning of the optical parameters with respect to the symmetric layout to be performed. Hence, this introduces a useful additional degree of freedom in the design of the proposed combined-function cell. 
\clearpage
\bibliographystyle{unsrt}
\bibliography{mybibliography}
\end{document}